\documentclass[12pt,english]{svjour3}
\usepackage{mathpazo}
\usepackage[T1]{fontenc}
\usepackage[utf8]{inputenc}
\usepackage{geometry}
\usepackage{babel}
\usepackage{array}
\usepackage{textcomp}
\usepackage{url}
\usepackage{multirow}
\usepackage{graphicx}
\usepackage{esint}
\usepackage[authoryear]{natbib}
\usepackage{subscript}
\usepackage[unicode=true,pdfusetitle,
 bookmarks=true,bookmarksnumbered=false,bookmarksopen=false,
 breaklinks=true,pdfborder={0 0 0},pdfborderstyle={},backref=false,colorlinks=false]
 {hyperref}

\makeatletter

\providecommand{\tabularnewline}{\\}

\makeatother

\begin{document}
\title{CMIP6 GCM ensemble members versus global surface temperatures}
\author{Nicola Scafetta\textsuperscript{}}
\institute{Department of Earth Sciences, Environment and Georesources, University
of Naples Federico II, Complesso Universitario di Monte S. Angelo,
via Cinthia, 21, 80126 Naples, Italy. Email: nicola.scafetta@unina.it}
\maketitle
\begin{abstract}
The Coupled Model Intercomparison Project (phase 6) (CMIP6) global
circulation models (GCMs) predict equilibrium climate sensitivity
(ECS) values ranging between 1.8${^\circ}$C and 5.7${^\circ}$C.
To narrow this range, we group 38 GCMs into low, medium and high ECS
subgroups and test their accuracy and precision in hindcasting the
mean global surface warming observed from 1980-1990 to 2011-2021 in
the ERA5-T2m, HadCRUT5, GISTEMP v4, and NOAAGlobTemp v5 global surface
temperature records. We also compare the GCM hindcasts to the satellite-based
UAH-MSU v6 lower troposphere global temperature record. We use 143
GCM ensemble averaged simulations under four slightly different forcing
conditions, 688 GCM member simulations, and Monte Carlo modeling of
the internal variability of the GCMs under three different model accuracy
requirements. We found that the medium and high-ECS GCMs run too hot
up to over 95\% and 97\% of cases, respectively. The low ECS GCM group
agrees best with the warming values obtained from the surface temperature
records, ranging between 0.52${^\circ}$C and 0.58${^\circ}$C. However,
when comparing the observed and GCM hindcasted warming on land and
ocean regions, the surface-based temperature records appear to exhibit a
significant warming bias. Furthermore, if the satellite-based UAH-MSU-lt
record is accurate, actual surface warming from 1980 to 2021 may have
been around 0.40${^\circ}$C (or less), that is up to about 30\% less
than what is reported by the surface-based temperature records. The
latter situation implies that even the low-ECS models would have produced
excessive warming from 1980 to 2021. These results suggest that the
actual ECS may be relatively low, i.e. lower than 3${^\circ}$C or
even less than 2${^\circ}$C if the 1980-2021 global surface temperature
records contain spurious warming, as some alternative studies have
already suggested. Therefore, the projected global climate warming
over the next few decades could be moderate and probably not particularly
alarming.
\end{abstract}

\subsubsection*{Keywords: CMIP6 climate models; temperature records; equilibrium
climate sensitivity; global warming; model validation and testing}

\subsubsection*{Cite this paper as: Scafetta, N., 2022. CMIP6 GCM ensemble members
versus global surface temperatures. Climate Dynamics. https://doi.org/10.1007/s00382-022-06493-w}

\section{Introduction}

The Coupled Model Intercomparison Project (phase 6) (CMIP6) collects
several simulations of global climate models (GCM) currently used
to interpret past and future climate changes \citep{Eyring,IPCC2021}.
However, these GCMs calculate equilibrium climate sensitivity (ECS)
values ranging from 1.8${^\circ}$C to 5.7${^\circ}$C \citep{IPCC2021}.
The ECS is the most important climatic parameter as it measures the
long-term increase in air temperature near the surface that should
result from an increase in radiative forcing of approximately 3.8
W/m\textsuperscript{2}, which corresponds to a doubling of the atmospheric
CO\textsubscript{2} concentration from 280 ppm (which is defined
as the preindustrial level) to 560 ppm. The uncertainty of the ECS
is highly problematic as it indicates that the climate system is still
poorly understood and modeled. Consequently, also the extent of future
climate change is rather uncertain as the impact of anthropogenic
CO\textsubscript{2} emissions on the climate cannot yet be adequately
quantified \citep[cf.][]{Knutti}.

The uncertainty of the ECS stems from the fact that various climate
feedback mechanisms -- in particular water vapor and cloud cover
-- are still too little known and modeled, as already found sixty
years ago by \citet{M=0000F6ller1963}. In the absence of climate
feedback mechanisms, the Stefan-Boltzmann law for blackbodies predicts
that a doubling of the atmospheric CO\textsubscript{2} concentration
could cause an increase in global surface temperature of about 1${^\circ}$C.
Therefore, only  strong positive climate feedbacks could significantly
increase the ECS above such a value, but their existence is still
debated.

Constraining the ECS value is an urgent task of climatology. In fact,
at least two-thirds of the CMIP6 GCMs could be severely defective.
For example, by grouping models into low ($1.5<ECS\leq3.0$ ${^\circ}$C),
medium ($3.0<ECS\leq4.5$ ${^\circ}$C) and high ($4.5<ECS\leq6.0$
${^\circ}$C) sensitivity values, if, say, the actual ECS is less
than 3°C, the GCMs with $ECS>3$ ${^\circ}$C should be ignored. Therefore,
it is very important that detailed evaluations of the models are carried
out in order to determine if, where and how the models should improve
both on a global scale -- as proposed, for example, in this work
-- and on regional scales, as done in numerous other studies \citep[e.g.:][and many others]{Heo,Seo}.

Constraining ECS also has important policy implications because the
expected warming for the 21\textsuperscript{st} century depends on
the value of the model's ECS \citep{Grose,Scafetta}: the higher the
ECS, the greater the expected warming due to GHG emissions. For example,
\citet{Huntingford} found that the wide ECS range of CMIP6 GCMs implies
that at thermal equilibrium the global surface temperature could warm
up between 1.0${^\circ}$C and 3.3${^\circ}$C above the pre-industrial
period (1850-1900) even if anthropocentric emissions cease today.

Scientists already wondered whether a strong response to greenhouse
gases could be realistic \citep{Voosen}. Indeed, high ECS CMIP6 models
have already been found to perform poorly \citep[e.g.:][]{Ribes,Scafetta,Tokarska,Zhu}
while the medium and even the low ECS models are being carefully evaluated.

For example, \citet{Nijsse} derived that the most likely ECS interval
should be 1.9-3.4${^\circ}$C while alternative studies, often empirical
based, have suggested that the actual ECS could be even lower, probably
between 1${^\circ}$C and 2.5${^\circ}$C \citep[e.g.:][]{Lewis,Lindzen,Scafetta2013,Stefani,Happer}.
Most GCMs seem to overestimate the observed surface warming since
1980 \citep{ScafettaMDPI2021,Scafetta} and also that observed in
the global \citep{McKitrick(2020)} and tropical troposphere \citep{Mitchell},
in particular at its top (200-300 hPa) where the CMIP6 GCMs predict
an unobserved hot spot \citep{McKitrick}. A similar situation also
occurred with the previous CMIP3 and CMIP5 GCMs \citep{Fu,Scafetta(2012a),Scafetta2013}.
Actually, as \citet{Knutti} acknowledged, there is a dichotomy between
the observed and modeled ECS as GCMs tend to favor sensitivity values
at the top of the probable range, while several studies based on instrumentally
recorded warming and some from paleoclimate favor values in the lower
part of the range. Therefore, not only the models with high ECS, but
also those with medium ECS should be and are being seriously questioned.

 \citet{Scafetta2021,Scafetta} showed that the performance of the
GCMs improves as their ECS decreases and, in any case, the low ECS
GCMs appear to be the best performing models. However, even low-ECS
GCMs need further evaluation because biases in some regions (e.g.
on land) could be offset by opposite biases in other regions (e.g.
on  ocean). Furthermore, serious uncertainties remain in the solar
forcing and in the temperature records themselves \citep{Connolly,DAleo(2016)}.
These uncertainties  question the warming trend reported by the available
climate records and, directly or indirectly, the models themselves.
Finally, climate systems seem to be regulated by various natural oscillations
from the decadal to the millennial scales, which the GCMs are unable
to reproduce, the presence of which would also imply low ECS values,
probably between 1 and 2${^\circ}$C \citep{Scafetta(2012a),Scafetta2013,Scafetta2021c}.

Focusing on the performance of the CMIP6 GCMs, \citet{Scafetta} proposed
that the probable ECS range could be constrained by  statistical investigation
to find which GCM group -- low, medium or high ECS -- best reproduces
the observed global surface warming between the 1980-1990 and 2011-2021
as reported by ERA5-T2m \citep{Hersbach,Simmons}. The period 1980-2021
was chosen because it is optimally covered by all available climatic
temperature records. \citet{Scafetta} analyzed the ``average''
simulations provided by the Koninklijk Nederlands Meteorologisch Instituut
(KNMI) Climate Explorer \citep{Oldenborgh} of 38 CMIP6 GCMs with
three shared socioeconomic pathways (SSP) emission scenarios, which
also counted for a partial evaluation of the internal variability
of the models. The low ECS GCM group was found to be perfectly compatible,
at least on a global scale, with the 2011-2021 warming relating to
the 1980-1990 period. Conversely, both GCM groups with medium and
high ECS showed too high warming trends.

A possible objection to the analysis proposed in \citet{Scafetta}
is that temperature records should be compared with actual members
of the CMIP6 GCM ensemble instead of their ensemble averages because
the unforced internal variability of the models produces different
results due to uncertainties in the initial conditions as well as
in the internal parameters of the models. This problem will be addressed
in this paper considering that:
\begin{enumerate}
\item physical models, including the GCMs, should be accurate and precise
(see Appendix B);
\item there are still open issues regarding the reliability of the available
global surface temperature records.
\end{enumerate}
In fact, theoretical models must reproduce observations within a reasonably
small error. In our case, it should be evident that the poor precision
of a GCM cannot be used as a pretext to justify its poor accuracy.
For example, a low-precision model could produce a very wide range
of different hindcasts due to its internal variability. In this situation,
even if some of its hindcasts fit the observations, the result should
still be considered unsatisfactory if the mean of the GCM set diverges
too much from the actual data. Similarly, if an ECS GCM group produces
a set of hindcasts that too sparsely encompass the observations, the
ECS values that characterize that group should be considered unrealistic
even though some of the models in the same group might perform better
than others. In general, the accuracy, precision and ECS category
of the GCMs must be evaluated simultaneously.

Furthermore, surface-based temperature records appear to exhibit non-climatic
warming biases due to poorly corrected urban heats or other local
surface phenomena \citep[e.g.:][]{Connolly,DAleo(2016),Scafetta2021}.
To account for this problem, the satellite temperature measurements
of the  lower troposphere using microwave resonance units (MSU) proposed
by the U. of Alabama Huntsville (UAH-MSU-lt v6) \citep{Spencer} will
also be analyzed.

UAH-MSU-lt is the temperature record that features the lowest global
warming trend (about +0.13 °C/decade) from 1980 to 2021 among all
available global temperature records. According to GCM simulations,
the troposphere is expected to warm up faster than the surface (up
to a factor of 3) because greenhouse gases are expected to warm the
atmosphere first \citep{Mitchell}. Consequently, the global warming
trend of the troposphere estimated from satellite measurements should
be further reduced to simulate the global warming trend at the surface.
Here, these corrections are ignored and UAH-MSU-lt is assumed to represent
the possible lowest limit for the global warming trend of the surface.
Therefore, comparison with this satellite temperature record could
help assess the presence of non-climatic warming bias in the surface
temperature records, particularly on land where large contaminated
areas appear to exist \citep[cf.][]{Ouyang2019,Scafetta2021}.

Indeed, preliminary analyzes have shown that the land  seems to have
warmed too much and too quickly compared to the ocean \citep{Scafetta2021}.
\citet{Connolly} used data from rural stations only and showed that
the warming of the Northern Hemisphere's land surface should be  significantly
lower than what reported by the available surface-based temperature
records based on both rural and urban stations. \citet{Watts(2022)}
examined the quality of the U.S. temperature stations from which official
temperature records are obtained and concluded that approximately
96\% of them could not meet the National Oceanic and Atmospheric Administration
(NOAA) requirements for ``acceptable placement'' because they could
be significantly contaminated by different heat sources. In general,
the surface temperature records and the homogenization algorithms
used to adjust them present several problems that may have exaggerated
the warming. Thus, the integrity of the available global surface temperature
records and, therefore, the ability to correctly determine the global
warming trend of the 20\textsuperscript{th} and 21\textsuperscript{st}
century should be questioned as well \citep{Connolly,DAleo(2016)}.

There is a different MSU record \citep{Mears}, which shows warming
trends that is more compatible with those presented by the surface-based
temperature records. However, this alternative satellite-based record
is not analyzed here because it would overlap the results of the surface-based
temperature records. In any case, adopting it in the present study
may not be optimal because it only covers the latitude range from
70.0${^\circ}$S to 82.5${^\circ}$N and because it appears to perform
worse than UAH-MSU-lt that better agrees with the radiosonde temperature
database \citep{Christy2018}.

Here, we significantly expand the analysis presented by \citet{Scafetta}
by testing 143 GCM average simulations  and all 688 GCM member simulations
available on the KNMI website against four surface-based global temperature
records (ERA5-T2m, HadCRUT5, GISTEMP v4, NOAAGlobTemp v5) and the
UAH-MSU-lt v6 satellite-based record. Since we wish to narrow the
ECS range, we again group the models into three classes corresponding
to low, medium and high ECS values, as proposed in \citet{Scafetta}.
ECS GCM groups that produce systematically biased trends (e.g. too
hot or too cold relative to the observed temperatures) should be questioned
and not used for policy even though some simulations may appear to
reproduce the observations. Finally, we compare the GCM hindcasts
with observed land and ocean warming values to determine whether the
surface-based records could be regionally biased and whether the ECS
should be further constrained towards lower values.

\begin{figure}[!t]
\centering{}\includegraphics[width=1\textwidth]{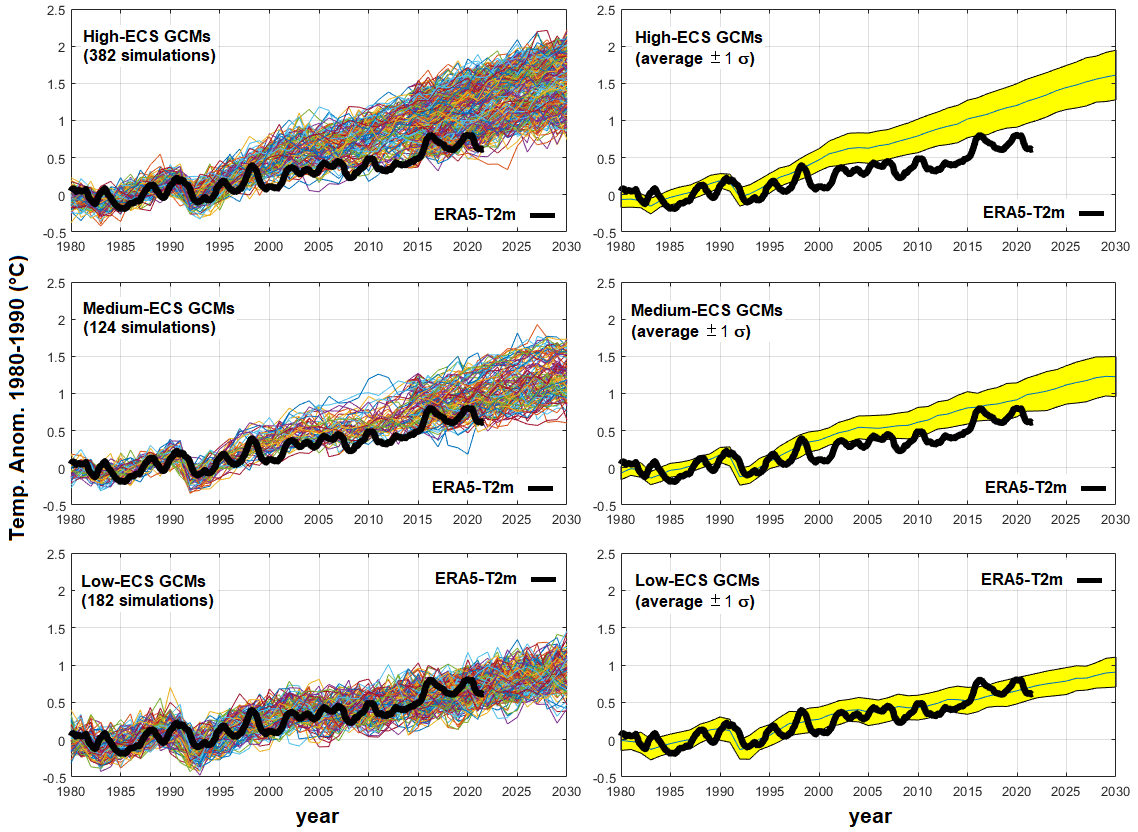}\caption{(Left) GCM global surface temperature simulations (colored curves)
and (right) $\pm1\sigma$ GCM global surface temperature ensembles
(yellow area) versus the ERA5-T2m record (black, 12-month moving average).}
\label{fig1}
\end{figure}

\begin{figure}[!t]
\centering{}\includegraphics[width=1\textwidth]{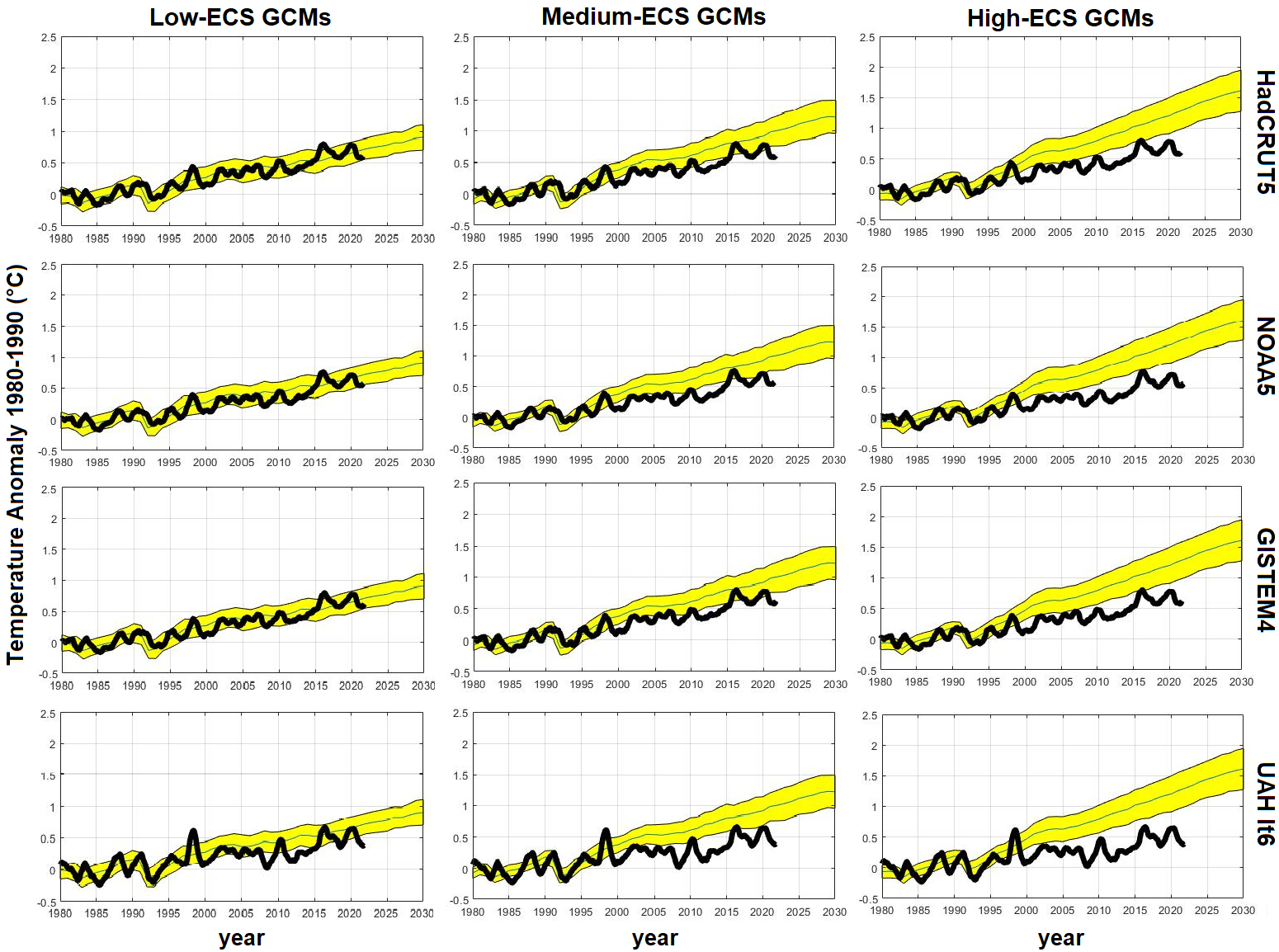}\caption{GCM global surface temperature ensembles (yellow area, $\pm1\sigma$)
versus HadCRUT5, GISTEMP v4, NOAAGlobTemp v5, and UAH-MSU-lt v6 temperature
records (black, 12-month moving average).}
\label{fig2}
\end{figure}

\section{Data and methods}

We analyze the monthly reanalysis field near-surface air temperature
(ERA5-T2m) record from 1980 to 2021 \citep{Hersbach,Simmons}. We
repeat the same analysis using the HadCRUT5 \citep{Morice}, GISTEMP
v4 \citep{Lenssen}, and NOAAGlobalTemp v5 \citep{Zhang} global surface
temperature records. Some of these records, however, may not cover
the entire surface of the globe from 1980 to 2021. There are other
global surface temperature records such as those proposed by the Japanese
Meteorological Agency \citep[JMA,][]{Ishihara} and by the Berkeley
Earth group \citep[BE,][]{Rohde}, which will also be discussed briefly.
For completeness, as explained in the Introduction, we add a comparison
with the UAH-MSU-lt v6 temperature measurements \citep{Spencer}.

We also analyze all 143 ``average'' surface air temperature (tas)
records and all 688 ensemble member records from 38 different CMIP6
GCMs downloadable from KNMI Climate Explorer. These simulations were
produced using historical forcings (1850-2014) further extended up
to 2100 with four different SSP scenarios: SSP1-2.6 (low GHG emissions),
SSP2-4.5 (intermediate GHG emissions), SSP3-7.0 (high GHG emissions
) and SSP5-8.5 (very high greenhouse gas emissions) \citep{IPCC2021}.
These four scenarios are nearly indistinguishable until 2021. Thus,
from 1850 to 2021 the four simulation sets can be considered independent
assessments of the same models under nearly identical forcing conditions,
which also helps to assess in first approximation the internal variability
of the models.

The 1980-2021 period was chosen to better evaluate the performance
of the CMIP6 GCMs. This period is optimally covered by numerous climatic
temperature records including those based on satellite measurements
that are alternative to those based on land and oceanic measurements
that could be affected by various non-climatic biases, which are difficult
to eliminate \citep{DAleo(2016),Watts(2022)}. In fact, going back
in time from 1980 to 1850, the temperature records are affected by
ever-larger uncertainties and uncovered areas, which makes evaluating
the CMIP6 models even more difficult. A possible advantage of the
present study is that the previous studies   evaluating the performance
of the CMIP6 models  attempted to constrain the ECS by comparing GCM
simulations only with surface climate records from 1850 to 2020 \citep{Ribes}
or from 1981 to 2014 \citep{Tokarska}, or even using uncertain paleoclimate
records \citep{Zhu} and concluded that only high-ECS models ($ECS>4.5$
${^\circ}$C) could be excluded. However, there are open questions
as to whether cooling adjustments applied to different Earth surface
temperature records from 1850 to 1980 are justified \citep{DAleo(2016)}
and whether in more recent periods the global surface climate records
are affected by non-climatic warming biases \citep{Connolly,Scafetta2021}.
These biases  could have exaggerated the 20\textsuperscript{th} century
warming trend and incorrectly provided support for the medium-ECS
GCMs.

The 1980-2021 warming for each record is calculated by evaluating
the 2011-2021 average temperature anomaly relative to the 1980-1990
period. 11-year intervals are used to bypass biases due to interannual
fluctuations such as those related to ENSO and the 11-year solar cycle.
Then, we apply standard statistical tests to decide if and how the
observed warming values for each of the temperature records are reproduced
by the three ECS GCM groups.

The ERA5-T2m global surface temperature average warming from 1980-1990
to 2011-2021 is estimated to be:

\begin{equation}
\Delta T_{mean}=0.578{^\circ}C.\label{eq:1}
\end{equation}
The other temperature records give: HadCRUT5 (infilled data), $\Delta T_{mean}=0.581$${^\circ}$C;
GISTEMP v4, $\Delta T_{mean}=0.570$${^\circ}$C; NOAAGlobalTemp v5,
$\Delta T_{mean}=0.523$${^\circ}$C. HadCRUT5, GISTEMP, and ERA5-T2m
give nearly identical warmings. We also observe that HadCRUT5 (non-infilled
data) gives $\Delta T_{mean}=0.549$${^\circ}$C and HadCRUT4 gives
$\Delta T_{mean}=0.521$${^\circ}$C. BE gives $\Delta T_{mean}=0.591$${^\circ}$C
and JMA gives $\Delta T_{mean}=0.557$${^\circ}$C, which do not differ
much from the above estimates. Thus, from 1980 to 2021, the available
surface-based global temperature records measure that the global surface
warming from 1980-1990 to 2011-2021 has been between 0.52${^\circ}$C
and 0.59${^\circ}$C, or approximately between 0.50${^\circ}$C and
0.60${^\circ}$C, with an average of 0.56${^\circ}$C. In contrast,
the satellite-based UAH-MSU-lt v6 temperature record gives $\Delta T_{mean}=0.402$${^\circ}$C,
suggesting that 2011-2021 actual warming may have been even less than
0.40${^\circ}$C because, as explained in the introduction, according
to the GCMs the temperature trend of the troposphere should be scaled
down to make it compatible with the surface warming trend.

For the temperature records, since 1980 the error of the average over
an 11-year period can be estimated to be very small, $\bar{\sigma}_{95\%}\approx0.01$${^\circ}$C
(see Appendix A), which represents about 2\% of the warming from 1980-1990
to 2011-2021, and is less than the differences between the various
temperature records.

As explained in the Introduction, the proposed analysis groups the
CMIP6 GCMs into three subsets characterized by low ($1.5<ECS\leq3.0$${^\circ}$C),
medium ($3.0<ECS\leq4.5$ ${^\circ}$C) and high ($4.5<ECS\leq6.0$
${^\circ}$C) sensitivity values. This choice is based on the following
heuristic considerations. In fact, the \citet{IPCC2013} estimated
that the ECS had to have a ``likely'' range of 1.5 -- 4.5${^\circ}$C.
Also \citet{Wigley} suggested the same interval although the best-fit
sensitivity was found to be 2.5°C. This range can be heuristically
divided into at least two equal parts: $1.5<ECS\leq3.0$ ${^\circ}$C
and $3.0<ECS\leq4.5$ ${^\circ}$C. In 2013, the CMIP5 GCMs were used.
However, the \citet{IPCC2021} adopted the CMIP6 GCMs that extended
the ECS range up to 6${^\circ}$C so that an equally large third range,
$4.5<ECS\leq6.0$ ${^\circ}$C, could be added to the previous two.
\citet{Zelinka} explained that the causes of the increased climate
sensitivity in the CMIP6 models were due to stronger positive cloud
feedbacks due to decreased extratropical cloud cover and albedo that,
however, might be questionable.

Therefore, the interval $1.5<ECS\leq3.0$ ${^\circ}$C  collects the
GCMs with ECS values most consistent with different empirical results,
as discussed in the Introduction; the interval $3.0<ECS\leq4.5$ ${^\circ}$C
 collects the other GCMs that also the \citet{IPCC2013} would have
considered acceptable; finally, the interval $4.5<ECS\leq6.0$ ${^\circ}$C
collects the GCMs included in the \citet{IPCC2021} but which in 2013
the IPCC itself considered to predict an unlikely high ECS.

\begin{figure}[!t]
\centering{}\includegraphics[width=1\textwidth]{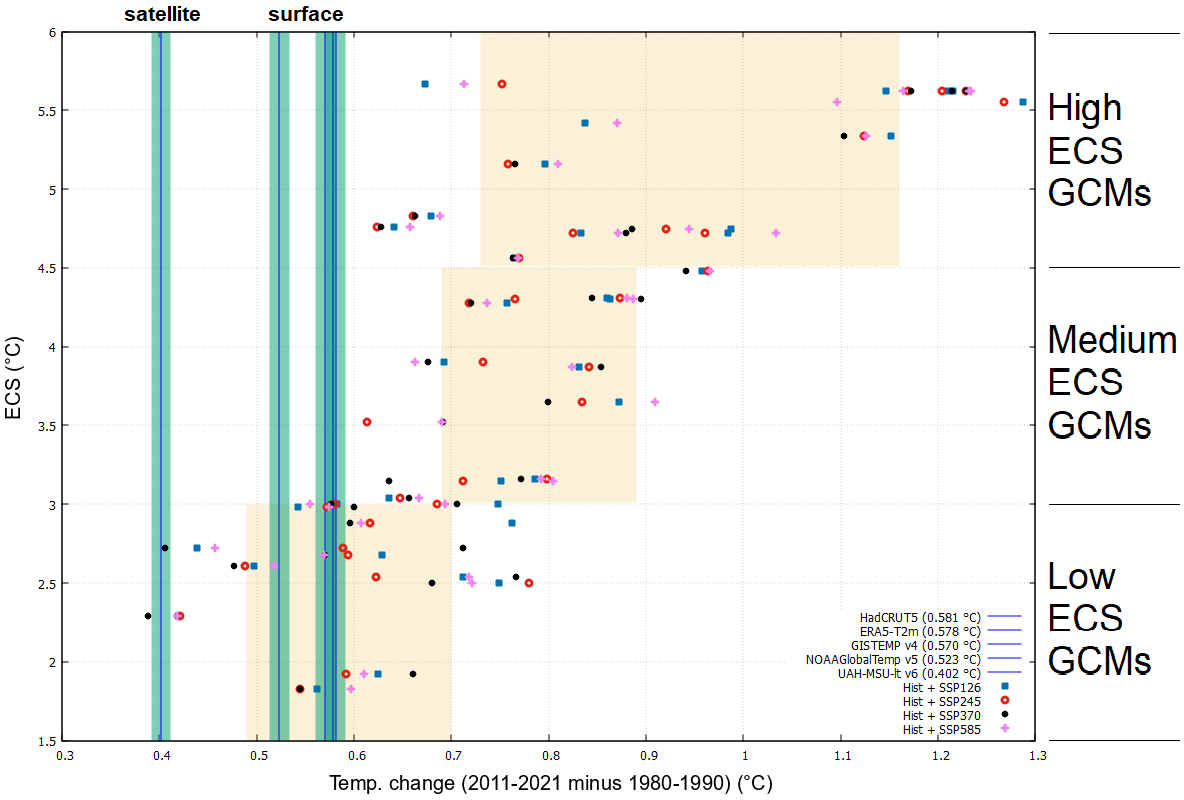}\caption{Average temperature changes (2011-2021 minus 1980-1990) hindcasted
by 38 CMIP6 GCMs mean simulations. The blue vertical lines represent
the temperature change measured by HadCRUT5, ERA5-T2m, GISTEMP v4,
NOAAGlobTemp v5, and UAH-MSU-lt v6 temperature records, respectively,
with their 95\% confidence interval. The three yellow boxes represent
the $\pm1\sigma$ dispersion of the data referring to the low ($1.5<ECS\protect\leq3.0$
${^\circ}$C), medium ($3.0<ECS\protect\leq4.5$ ${^\circ}$C) and
high ($4.5<ECS\protect\leq6.0$ ${^\circ}$C) ECS GCMs. See Table
\ref{Tab1}.}
\label{fig3}
\end{figure}

\begin{figure}[!t]
\centering{}\includegraphics[width=1\textwidth]{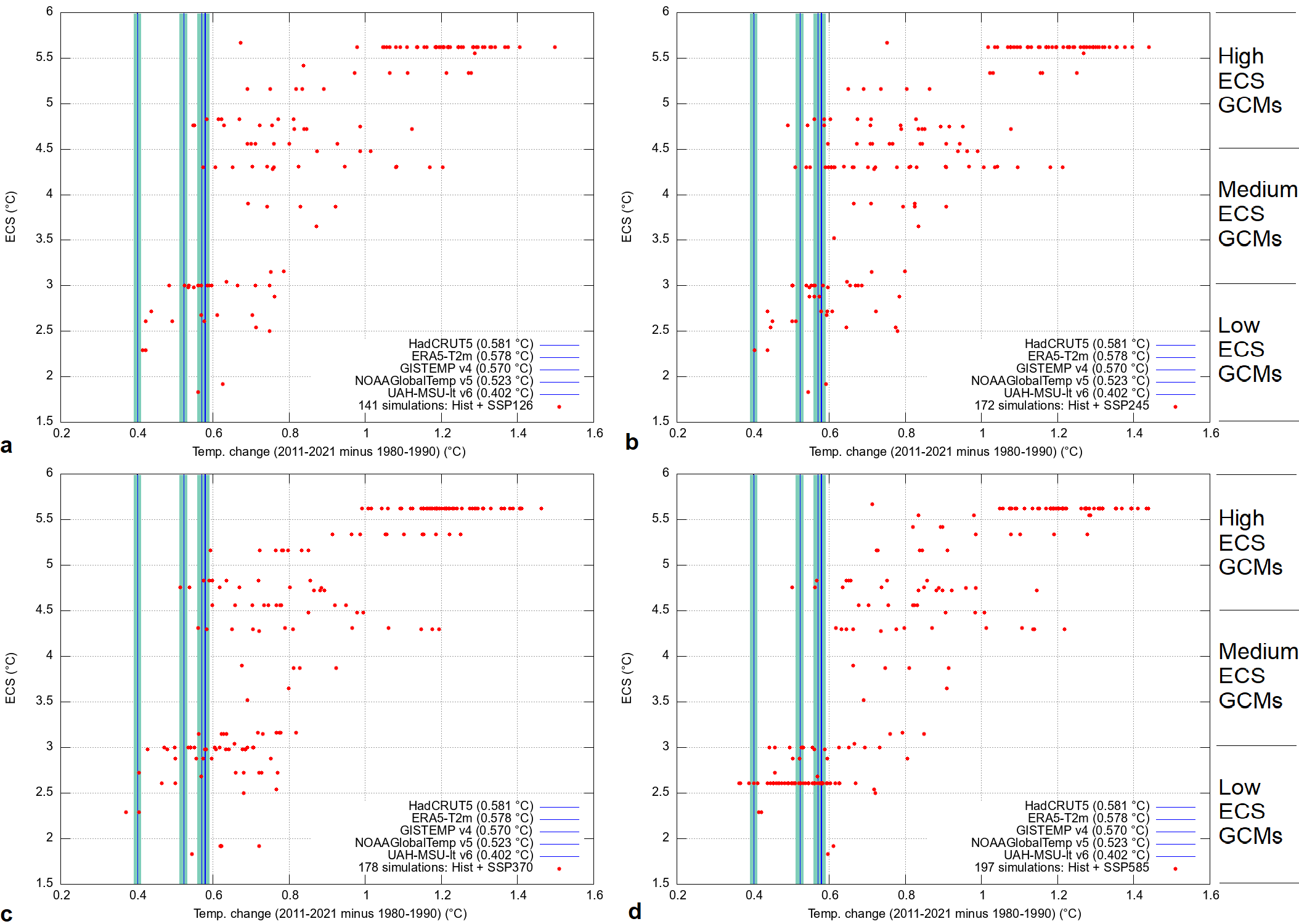}\caption{Temperature change (2011-2021 minus 1980-1990) hindcasted by the full
range of model simulations (red dots). The vertical lines represents
the global surface warming from 1980-1990 to 2011-2021 reported by
HadCRUT5, ERA5-T2m, GISTEMP v4, NOAAGlobTemp v5, and UAH-MSU-lt v6
temperature records, respectively, with their 95\% confidence interval.
Each of the four panel represents a set of forcing: (a) Hist+SSP1-2.6,
(b) Hist+SSP2-4.5, (c) Hist+SSP3-7.0 and (d) Hist+SSP5-8.5.}
\label{fig4}
\end{figure}

\begin{table}
\begin{centering}
\begin{tabular}{llcccccl||l}
﻿ECS GROUP & ﻿GCM & ECS (${^\circ}$C) & SSP1-2.6 (${^\circ}$C) & SSP2-4.5 (${^\circ}$C) & SSP3-7.0 (${^\circ}$C) & SSP5-8.5 (${^\circ}$C) & \multicolumn{2}{c}{$mean\pm1\sigma$ (${^\circ}$C)}\tabularnewline
\hline 
\multirow{15}{*}{High ECS} & CIESM & 5.67 & 0.67 & 0.75 &  & 0.71 & \multicolumn{2}{c}{$0.71\pm0.04$}\tabularnewline
 & CanESM5-CanOE-p2 & 5.62 & 1.15 & 1.17 & 1.17 & 1.16 & \multicolumn{2}{c}{$1.16\pm0.01$}\tabularnewline
 & CanESM5-p1 & 5.62 & 1.21 & 1.20 & 1.21 & 1.23 & \multicolumn{2}{c}{$1.22\pm0.01$}\tabularnewline
 & CanESM5-p2 & 5.62 & 1.22 & 1.23 & 1.23 & 1.23 & \multicolumn{2}{c}{$1.23\pm0.01$}\tabularnewline
 & HadGEM3-GC31-LL-f3 & 5.55 & 1.29 & 1.27 &  & 1.10 & \multicolumn{2}{c}{$1.22\pm0.11$}\tabularnewline
 & HadGEM3-GC31-MM-f3 & 5.42 & 0.84 &  &  & 0.87 & \multicolumn{2}{c}{$0.85\pm0.02$}\tabularnewline
 & UKESM1-0-LL-f2 & 5.34 & 1.15 & 1.12 & 1.10 & 1.13 & \multicolumn{2}{c}{$1.13\pm0.02$}\tabularnewline
 & CESM2 & 5.16 & 0.80 & 0.76 & 0.77 & 0.81 & \multicolumn{2}{c}{$0.78\pm0.02$}\tabularnewline
 & CNRM-CM6-1-f2 & 4.83 & 0.68 & 0.66 & 0.66 & 0.69 & \multicolumn{2}{c}{$0.67\pm0.01$}\tabularnewline
 & CNRM-ESM2-1-f2 & 4.76 & 0.64 & 0.62 & 0.63 & 0.66 & \multicolumn{2}{c}{$0.64\pm0.02$}\tabularnewline
 & CESM2-WACCM & 4.75 & 0.99 & 0.92 & 0.89 & 0.94 & \multicolumn{2}{c}{$0.93\pm0.04$}\tabularnewline
 & ACCESS-CM2 & 4.72 & 0.83 & 0.82 & 0.88 & 0.87 & \multicolumn{2}{c}{$0.85\pm0.03$}\tabularnewline
 & NESM3 & 4.72 & 0.98 & 0.96 &  & 1.03 & \multicolumn{2}{c}{$0.99\pm0.04$}\tabularnewline
 & IPSL-CM6A-LR & 4.56 & 0.76 & 0.77 & 0.76 & 0.77 & \multicolumn{2}{c}{$0.77\pm0.01$}\tabularnewline
 & $mean\pm1\sigma$ &  & $0.94\pm0.23$ & $0.94\pm0.23$ & $0.93\pm0.23$ & $0.94\pm0.20$ & \multicolumn{2}{c}{$0.94\pm0.23$}\tabularnewline
\hline 
\multirow{12}{*}{Medium ECS} & KACE-1-0-G & 4.48 & 0.96 & 0.96 & 0.94 & 0.97 & \multicolumn{2}{c}{$0.96\pm0.01$}\tabularnewline
 & EC-Earth3-Veg & 4.31 & 0.86 & 0.87 & 0.84 & 0.88 & \multicolumn{2}{c}{$0.86\pm0.02$}\tabularnewline
 & EC-Earth3 & 4.3 & 0.86 & 0.76 & 0.90 & 0.89 & \multicolumn{2}{c}{$0.85\pm0.06$}\tabularnewline
 & CNRM-CM6-1-HR-f2 & 4.28 & 0.76 & 0.72 & 0.72 & 0.74 & \multicolumn{2}{c}{$0.73\pm0.02$}\tabularnewline
 & GFDL-ESM4 & 3.9 & 0.69 & 0.73 & 0.68 & 0.66 & \multicolumn{2}{c}{$0.69\pm0.03$}\tabularnewline
 & ACCESS-ESM1-5 & 3.87 & 0.83 & 0.84 & 0.85 & 0.82 & \multicolumn{2}{c}{$0.84\pm0.01$}\tabularnewline
 & MCM-UA-1-0 & 3.65 & 0.87 & 0.83 & 0.80 & 0.91 & \multicolumn{2}{c}{$0.85\pm0.05$}\tabularnewline
 & CMCC-CM2-SR5 & 3.52 &  & 0.61 & 0.69 & 0.69 & \multicolumn{2}{c}{$0.66\pm0.04$}\tabularnewline
 & AWI-CM-1-1-MR & 3.16 & 0.79 & 0.80 & 0.77 & 0.79 & \multicolumn{2}{c}{$0.79\pm0.01$}\tabularnewline
 & MRI-ESM2-0 & 3.15 & 0.75 & 0.71 & 0.64 & 0.80 & \multicolumn{2}{c}{$0.73\pm0.07$}\tabularnewline
 & BCC-CSM2-MR & 3.04 & 0.64 & 0.65 & 0.66 & 0.67 & \multicolumn{2}{c}{$0.65\pm0.01$}\tabularnewline
 & $mean\pm1\sigma$ &  & $0.80\pm0.10$ & $0.77\pm0.10$ & $0.77\pm0.10$ & $0.80\pm0.10$ & \multicolumn{2}{c}{$0.79\pm0.10$}\tabularnewline
\hline 
\multirow{14}{*}{Low ECS} & FGOALS-f3-L & 3 & 0.75 & 0.69 & 0.71 & 0.69 & \multicolumn{2}{c}{$0.71\pm0.03$}\tabularnewline
 & MPI-ESM1-2-LR & 3 & 0.58 & 0.58 & 0.58 & 0.56 & \multicolumn{2}{c}{$0.57\pm0.01$}\tabularnewline
 & MPI-ESM1-2-HR & 2.98 & 0.54 & 0.57 & 0.60 & 0.57 & \multicolumn{2}{c}{$0.57\pm0.02$}\tabularnewline
 & FGOALS-g3 & 2.88 & 0.76 & 0.62 & 0.60 & 0.61 & \multicolumn{2}{c}{$0.65\pm0.08$}\tabularnewline
 & GISS-E2-1-G-p1 & 2.72 &  &  & 0.71 &  & \multicolumn{2}{c}{$0.71$}\tabularnewline
 & GISS-E2-1-G-p3 & 2.72 & 0.44 & 0.59 & 0.41 & 0.46 & \multicolumn{2}{c}{$0.47\pm0.08$}\tabularnewline
 & MIROC-ES2L-f2 & 2.68 & 0.63 & 0.59 & 0.57 & 0.57 & \multicolumn{2}{c}{$0.59\pm0.03$}\tabularnewline
 & MIROC6 & 2.61 & 0.50 & 0.49 & 0.48 & 0.52 & \multicolumn{2}{c}{$0.50\pm0.02$}\tabularnewline
 & NorESM2-LM & 2.54 & 0.71 & 0.62 & 0.77 & 0.72 & \multicolumn{2}{c}{$0.70\pm0.06$}\tabularnewline
 & NorESM2-MM & 2.5 & 0.75 & 0.78 & 0.68 & 0.72 & \multicolumn{2}{c}{$0.73\pm0.04$}\tabularnewline
 & CAMS-CSM1-0 & 2.29 & 0.42 & 0.42 & 0.39 & 0.42 & \multicolumn{2}{c}{$0.41\pm0.02$}\tabularnewline
 & INM-CM5-0 & 1.92 & 0.62 & 0.59 & 0.66 & 0.61 & \multicolumn{2}{c}{$0.62\pm0.03$}\tabularnewline
 & INM-CM4-8 & 1.83 & 0.56 & 0.54 & 0.54 & 0.60 & \multicolumn{2}{c}{$0.56\pm0.02$}\tabularnewline
 & $mean\pm1\sigma$ &  & $0.61\pm0.12$ & $0.59\pm0.09$ & $0.59\pm0.12$ & $0.59\pm0.09$ & \multicolumn{2}{c}{$0.60\pm0.12$}\tabularnewline
\end{tabular}
\par\end{centering}
\caption{Warming from 1980-1990 to 2011-2021 for average simulations of 38
GCMs using Historical + SSP1-2.6, SSP2-4.5, SSP3-7.0, and SSP5-8.5
forcings. See Figure \ref{fig3}.}
\label{Tab1}
\end{table}

\begin{table}
\centering{}%
\begin{tabular}{cccccccc}
\hline 
ECS GROUP &  & SSP1-2.6 & SSP2-4.5 & SSP3-7.0 & SSP5-8.5 & total & \%\tabularnewline
 & total members & 141 & 172 & 178 & 197 & 688 & \tabularnewline
\hline 
\multirow{10}{*}{High ECS} & $\mathrm{Model>HadCRUT}$ & 86 & 93 & 97 & 93 & 369 & 97\%\tabularnewline
 & $\mathrm{Model>ERA5T2m}$ & 86 & 93 & 97 & 93 & 369 & 97\%\tabularnewline
 & $\mathrm{Model>GISTEMP}$ & 88 & 93 & 98 & 93 & 372 & 98\%\tabularnewline
 & $\mathrm{Model>NOAAGT}$ & 88 & 95 & 100 & 96 & 379 & 99.5\%\tabularnewline
 & $\mathrm{Model>UAHMSU}$ & 88 & 96 & 100 & 96 & 380 & 100\%\tabularnewline
\cline{2-8} \cline{3-8} \cline{4-8} \cline{5-8} \cline{6-8} \cline{7-8} \cline{8-8} 
 & $\mathrm{Model<HadCRUT}$ & 2 & 3 & 3 & 3 & 11 & 3\%\tabularnewline
 & $\mathrm{Model<ERA5T2m}$ & 2 & 3 & 3 & 3 & 11 & 3\%\tabularnewline
 & $\mathrm{Model<GISTEMP}$ & 0 & 3 & 2 & 3 & 8 & 2\%\tabularnewline
 & $\mathrm{Model<NOAAGT}$ & 0 & 1 & 0 & 0 & 1 & 0.5\%\tabularnewline
 & $\mathrm{Model<UAHMSU}$ & 0 & 0 & 0 & 0 & 0 & 0\%\tabularnewline
\hline 
\multirow{10}{*}{Medium ECS} & $\mathrm{Model>HadCRUT}$ & 25 & 39 & 29 & 26 & 119 & 94\%\tabularnewline
 & $\mathrm{Model>ERA5T2m}$ & 25 & 39 & 30 & 26 & 120 & 95\%\tabularnewline
 & $\mathrm{Model>GISTEMP}$ & 26 & 39 & 30 & 26 & 121 & 96\%\tabularnewline
 & $\mathrm{Model>NOAAGT}$ & 26 & 41 & 32 & 26 & 125 & 99\%\tabularnewline
 & $\mathrm{Model>UAHMSU}$ & 26 & 42 & 32 & 26 & 126 & 100\%\tabularnewline
\cline{2-8} \cline{3-8} \cline{4-8} \cline{5-8} \cline{6-8} \cline{7-8} \cline{8-8} 
 & $\mathrm{Model<HadCRUT}$ & 1 & 3 & 3 & 0 & 7 & 6\%\tabularnewline
 & $\mathrm{Model<ERA5T2m}$ & 1 & 3 & 2 & 0 & 6 & 5\%\tabularnewline
 & $\mathrm{Model<GISTEMP}$ & 0 & 3 & 2 & 0 & 5 & 4\%\tabularnewline
 & $\mathrm{Model<NOAAGT}$ & 0 & 1 & 0 & 0 & 1 & 1\%\tabularnewline
 & $\mathrm{Model<UAHMSU}$ & 0 & 0 & 0 & 0 & 0 & 0\%\tabularnewline
\hline 
\multirow{10}{*}{Low ECS} & $\mathrm{Model>HadCRUT}$ & 12 & 14 & 25 & 18 & 69 & 38\%\tabularnewline
 & $\mathrm{Model>ERA5T2m}$ & 12 & 15 & 26 & 19 & 72 & 40\%\tabularnewline
 & $\mathrm{Model>GISTEMP}$ & 13 & 16 & 28 & 24 & 81 & 45\%\tabularnewline
 & $\mathrm{Model>NOAAGT}$ & 21 & 26 & 35 & 39 & 121 & 66\%\tabularnewline
 & $\mathrm{Model>UAHMSU}$ & 27 & 33 & 43 & 71 & 174 & 96\%\tabularnewline
\cline{2-8} \cline{3-8} \cline{4-8} \cline{5-8} \cline{6-8} \cline{7-8} \cline{8-8} 
 & $\mathrm{Model<HadCRUT}$ & 15 & 20 & 21 & 57 & 113 & 62\%\tabularnewline
 & $\mathrm{Model<ERA5T2m}$ & 15 & 19 & 20 & 56 & 110 & 60\%\tabularnewline
 & $\mathrm{Model<GISTEMP}$ & 14 & 18 & 18 & 51 & 101 & 55\%\tabularnewline
 & $\mathrm{Model<NOAAGT}$ & 6 & 8 & 11 & 36 & 61 & 34\%\tabularnewline
 & $\mathrm{Model<UAHMSU}$ & 0 & 1 & 3 & 4 & 8 & 4\%\tabularnewline
\hline 
\end{tabular}\caption{Number of single GCM simulations reporting mean temperature changes
(2011-2021 minus 1980-1990) lower or higher than HadCRUT5, ERA5-T2m,
GISTEMP v4, NOAAGlobTemp v5 and UAH-MSU-lt v6, respectively.}
\label{tab2}
\end{table}

\section{Analysis of the CMIP6 GCM simulations}

Figure \ref{fig1} shows the GCM simulations (left) and their ensemble
$mean\pm1\sigma$ range (right) grouped according to the three GCM
ECS sets with respect to the ERA5-T2m global surface temperature record
(black, moving averages at 12 months). All records are temperature
anomalies relative to the period 1980-1990. Figure \ref{fig2} shows
a similar comparison with respect to the HadCRUT, GISTEMP, NOAAGlobTemp
and UAH-MSU-lt temperature records.

Both figures show that as the ECS increases, the global surface warming
predicted by the models also increases. However, only the low-ECS
GCM group can be considered perfectly consistent with the surface-based
global temperature records because it encloses them well within the
$\pm1$$\sigma$ GCM range (yellow area).

Figures \ref{fig1} and \ref{fig2} also show that, compared to the
satellite record, even the GCM group with low ECS seems to overestimate
the observed warming. In fact, even for the low ECS GCM group from
2011 to 2021 the UAH-MSU-lt record is not well enclosed within the
$\pm1\sigma$ model ensemble (yellow) area although a better agreement
is found in the period 2015- 2020. The latter was characterized by
the significant El Niño warming events of 2015-2016 and 2020 (Appendix
A, Figure \ref{figA1}). Therefore, the 2015-2020 warming for the
period 2000-2014 could also be temporary \citep{Scafetta2021c} and
not related to the warming hindcasted by the models because it is
clearly due to natural climatic fluctuations while the average warming
produced by the models is due to anthropogenic forcing. From 2015
to 2022, in fact, a slightly cooling trend is observed. From 2000
to 2014 the UAH-MSU-lt v6 record also clearly shows the so-called global
warming “hiatus” or “pause” \citep{IPCC2013}. This decade-long lack
of warming began to seriously question the GCMs, and various statistical
solutions were proposed to circumvent the problem by referring to
the fluctuations of the internal variability of the climate system
\citep[e.g.][]{Meehl}. Figures \ref{fig1} and \ref{fig2} also show that,
at the present, the \textquotedbl pause\textquotedbl{} appears missing
or attenuated in the latest versions of the surface-based  global temperature
records.

\subsection{Analysis of the GCM average simulations}

\citet{Scafetta} analyzed the average simulations of 38 GCMs using
the historical + SSP2-4.5, SSP3-7.0, and SSP5-8.5 radiative forcing
scenarios up to June 2021; the warming values for each model were
collected in the table there published. Figure \ref{fig3} graphically
shows the results of the same analysis, which was updated to the whole
year 2021 and  also included the SSP1-2.6 simulations, compared to
the temperature observations (green vertical lines). 143 average records
are analyzed. For each ECS GCM group the statistics provide (see Table
\ref{Tab1}):
\begin{itemize}
\item High-ECS GCMs (51 records): $\Delta T_{mean}=0.94\pm0.22$ ${^\circ}$C;
\item Medium-ECS GCMs (43 records): $\Delta T_{mean}=0.79\pm0.10$ ${^\circ}$C;
\item Low-ECS GCMs (49 records): $\Delta T_{mean}=0.59\pm0.10$ ${^\circ}$C.
\end{itemize}
\noindent The result confirms that the GCM group with low ECS is perfectly
compatible with the observed warming (Eq. \ref{eq:1}) within the
$\pm1$$\sigma$ range. In contrast, both GCM groups with medium and
high ECS show warming biases. Moreover, as \citet{Scafetta} already
observed, Figure \ref{fig3} also shows that none of the medium and
high ECS models predict an average warming of less than 0.6${^\circ}$C,
which is above the warming reported by all global temperature surface
records. This result suggests that models with $ECS>3$${^\circ}$C
should be questioned at the 95\% confidence level. Thus, by considering
only the GCM ensemble averages for the four SSPs, the real ECS should
be equal to or lower than 3${^\circ}$C.

\noindent However, Figure \ref{fig3} also shows that if the UAH-MSU-LT
record better reproduces the actual 2011-2021 warming, the GCM group
with low ECS would also be too hot because, out of 49 GCM ensemble
averages with low ECS, 48 cases (98\%) are warmer than 0.40${^\circ}$C.
The GCM that best agrees with the satellite record is CAMS-CSM1-0
whose  ECS is 2.29${^\circ}$C.

\begin{figure}[!t]
\centering{}\includegraphics[width=1\textwidth]{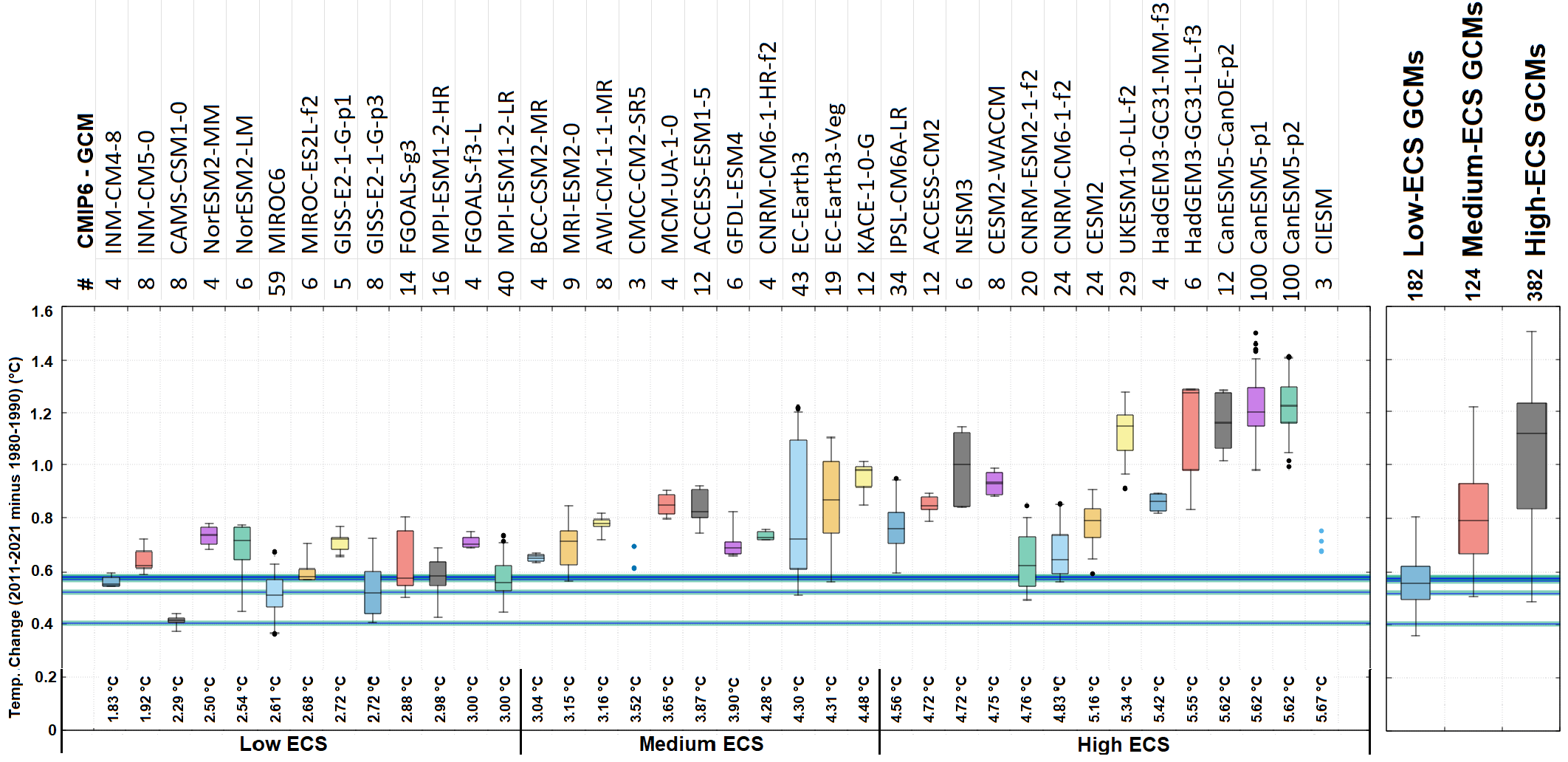}\caption{Boxplots of the CMIP6 ensemble members depicted in Figure \ref{fig4}
for each CMIP6 GCM; \# represents the number of the available simulations
for each GCM. The horizontal blue lines represents the global surface
warming from 1980-1990 to 2011-2021 reported by HadCRUT5, ERA5-T2m,
GISTEMP v4, NOAAGlobTemp v5, and UAH-MSU-lt v6 temperature records,
respectively. The whiskers extend from each end of the box for a range
up to 1.5 times the interquartile range.}
\label{fig5}
\end{figure}

\begin{figure}[!t]
\centering{}\includegraphics[width=1\textwidth]{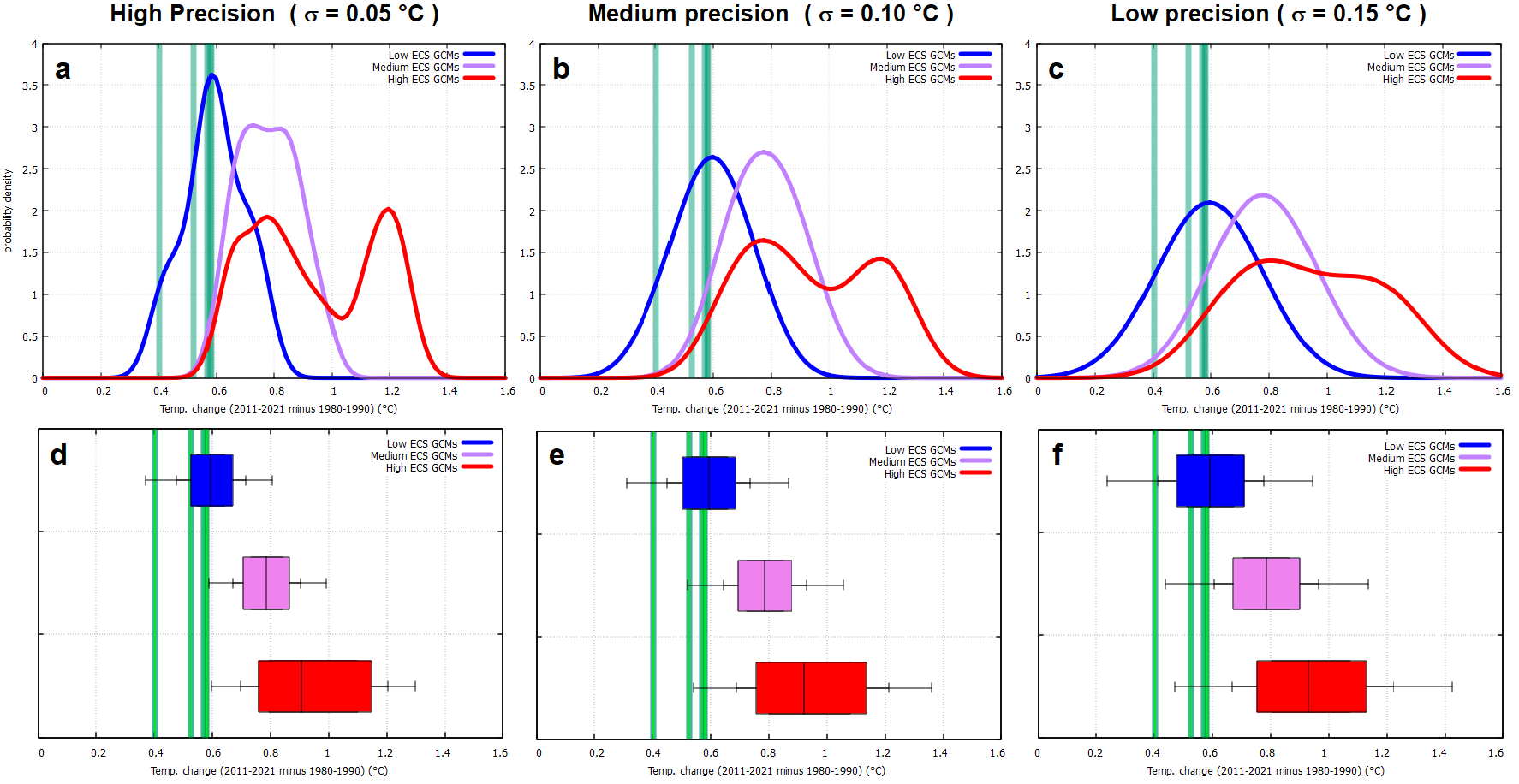}\caption{(a, b, c) GCM probability density functions relative to three model
precision requirements: $\sigma=0.05$${^\circ}$C, $\sigma=0.10$${^\circ}$C
and $\sigma=0.15$${^\circ}$C, which approximately correspond to
$\sigma_{95\%}=0.10$${^\circ}$C, $\sigma_{95\%}=0.20$${^\circ}$C
and $\sigma_{95\%}=0.30$${^\circ}$C. The green vertical lines represent
the global surface warming from 1980-1990 to 2011-2021 reported by
HadCRUT5, ERA5-T2m, GISTEMP v4, NOAAGlobTemp v5 and UAH-MSU-lt v6
temperature records. (d, e, f) Boxplots of the the probability density
functions depicted in panels a, b, c, respectively, using double whiskers
and boxes indicating the following probability ranges: 2.5\%, 16\%,
25\%, 50\%, 75\%, 84\% and 97.5\%.}
\label{fig6}
\end{figure}

\subsection{Analysis of the full range of the GCM ensemble members}

Figure \ref{fig4} shows in four panels the temperature variations
(2011-2021 minus 1980-1990) of the 688 simulations of GCM ensemble
members available per forcing set (Hist+SSP1-2.6, Hist+SSP2-4.5, Hist+SSP3-7.0
and Hist+SSP5-8.5; red dots) against the five temperature records
(vertical lines). The figure visually confirms that the vast majority
of ensemble member simulations produced by the GCM groups with medium
and high ECS run too hot relative to all five temperature records.

To examine how observed warming values are placed within the distributions
of possible GCM hindcasts for each of the three ECS groups, we count
how many member simulations record temperatures colder or warmer than
each of the five temperature records. Table \ref{tab2} reports the
results.

The analysis confirms that the low-ECS models produce results that
well enclose the 2011-2021 average temperatures obtained using the
surface temperature temperature records, which always fall within
the statistical interval $\pm1\sigma$ (corresponding to the 16-84\%
probability interval) of the distribution of the GCM hindcasts. In
contrast, 94-100\% and 97-100\% of hindcasts produced by the GCMs
with medium and high ECS are warmer than all five temperature records,
respectively. Therefore, also considering the full range of the available
CMIP6 GCM simulations, the GCMs with medium and high ECS run too hot.
Thus, the actual ECS should be equal to or less than 3${^\circ}$C.

However, 96\% of GCM simulations from the low-ECS GCM are warmer and
only 4\% cooler than the lower troposphere temperature record. So,
once again, we found that if UAH-MSU-lt better reproduces the actual
global warming from 1980-1990 to 2011-2021, the vast majority of the
low-ECS GCM ensemble members would also be found to run too hot.

\begin{table}
\centering{}%
\begin{tabular}{clccc}
\hline 
ECS GROUP &  & high precision & medium precision & low precision\tabularnewline
 &  & ( $\sigma_{H}=0.05$${^\circ}$C) & ( $\sigma_{M}=0.10$${^\circ}$C) & ( $\sigma_{L}=0.15$${^\circ}$C)\tabularnewline
\hline 
\multirow{10}{*}{High ECS} & $\mathrm{HadCRUT<Model}$ & 99\% & 95\% & 92\%\tabularnewline
 & $\mathrm{ERA5T2m<Model}$ & 99\% & 95\% & 92\%\tabularnewline
 & $\mathrm{GISTEMP<Model}$ & 99\% & 96\% & 93\%\tabularnewline
 & $\mathrm{NOAAGT<Model}$ & 100\% & 98\% & 95\%\tabularnewline
 & $\mathrm{UAHMSU<Model}$ & 100\% & 100\% & 99\%\tabularnewline
\cline{2-5} \cline{3-5} \cline{4-5} \cline{5-5} 
 & $\mathrm{HadCRUT>Model}$ & 1\% & 5\% & 8\%\tabularnewline
 & $\mathrm{ERA5T2m>Model}$ & 1\% & 5\% & 8\%\tabularnewline
 & $\mathrm{GISTEMP>Model}$ & 1\% & 4\% & 7\%\tabularnewline
 & $\mathrm{NOAAGT>Model}$ & 0\% & 2\% & 5\%\tabularnewline
 & $\mathrm{UAHMSU>Model}$ & 0\% & 0\% & 1\%\tabularnewline
\hline 
\multirow{10}{*}{Medium ECS} & $\mathrm{HadCRUT<Model}$ & 98\% & 93\% & 88\%\tabularnewline
 & $\mathrm{ERA5T2m<Model}$ & 98\% & 93\% & 88\%\tabularnewline
 & $\mathrm{GISTEMP<Model}$ & 99\% & 94\% & 88\%\tabularnewline
 & $\mathrm{NOAAGT<Model}$ & 100\% & 97\% & 93\%\tabularnewline
 & $\mathrm{UAHMSU<Model}$ & 100\% & 100\% & 99\%\tabularnewline
\cline{2-5} \cline{3-5} \cline{4-5} \cline{5-5} 
 & $\mathrm{HadCRUT>Model}$ & 2\% & 7\% & 12\%\tabularnewline
 & $\mathrm{ERA5T2m>Model}$ & 2\% & 7\% & 12\%\tabularnewline
 & $\mathrm{GISTEMP>Model}$ & 1\% & 6\% & 12\%\tabularnewline
 & $\mathrm{NOAAGT>Model}$ & 0\% & 3\% & 7\%\tabularnewline
 & $\mathrm{UAHMSU>Model}$ & 0\% & 0\% & 1\%\tabularnewline
\hline 
\multirow{10}{*}{Low ECS} & $\mathrm{HadCRUT<Model}$ & 55\% & 54\% & 53\%\tabularnewline
 & $\mathrm{ERA5T2m<Model}$ & 56\% & 55\% & 53\%\tabularnewline
 & $\mathrm{GISTEMP<Model}$ & 59\% & 57\% & 55\%\tabularnewline
 & $\mathrm{NOAAGT<Model}$ & 74\% & 69\% & 65\%\tabularnewline
 & $\mathrm{UAHMSU<Model}$ & 95\% & 91\% & 85\%\tabularnewline
\cline{2-5} \cline{3-5} \cline{4-5} \cline{5-5} 
 & $\mathrm{HadCRUT>Model}$ & 45\% & 46\% & 47\%\tabularnewline
 & $\mathrm{ERA5T2m>Model}$ & 44\% & 45\% & 47\%\tabularnewline
 & $\mathrm{GISTEMP>Model}$ & 41\% & 43\% & 45\%\tabularnewline
 & $\mathrm{NOAAGT>Model}$ & 26\% & 31\% & 35\%\tabularnewline
 & $\mathrm{UAHMSU>Model}$ & 5\% & 9\% & 15\%\tabularnewline
\hline 
\end{tabular}\caption{Probability $P_{\Delta T<GCMs}$ and $P_{\Delta T>GCMs}$ that the
2011-2021 warming hindcast from 1980-1990 to 2011-2021 for each ECS
GCM ensemble is warmer or colder, respectively, than HadCRUT5, ERA5-
T2m, GISTEMP v4, NOAAGlobTemp v5 and UAH-MSU-lt v6, respectively:
see Figure \ref{fig6}.}
\label{tab3}
\end{table}

\begin{table}
\centering{}%
\begin{tabular}{lcccc|cccc}
\hline 
 & \multicolumn{4}{c|}{Observations and GCM hindcasts} & \multicolumn{4}{c}{Modeled Global and Land Warmings}\tabularnewline
\hline 
﻿(80${^\circ}$S:80${^\circ}$N) & Total (${^\circ}$C) & Land (${^\circ}$C) & Ocean (${^\circ}$C) & ratio & Total (${^\circ}$C) & Land (${^\circ}$C) & Ocean (${^\circ}$C) & average ratio\tabularnewline
HadCRUT5 & 0.57 & 0.86 & 0.45 & 1.92 & 0.54 & 0.76 & 0.45 & 1.69\tabularnewline
ERA5-T2m & 0.57 & 0.92 & 0.43 & 2.14 & 0.51 & 0.73 & 0.43 & 1.69\tabularnewline
GISTEMP & 0.54 & 0.93 & 0.41 & 2.26 & 0.48 & 0.69 & 0.41 & 1.69\tabularnewline
NOAAGT & 0.53 & 0.91 & 0.38 & 2.38 & 0.45 & 0.64 & 0.38 & 1.69\tabularnewline
MSU-UAH & 0.40 & 0.53 & 0.34 & 1.54 & 0.41 & 0.57 & 0.34 & 1.69\tabularnewline
Low ECS GCMs & 0.58$\pm$0.10 & 0.84$\pm$0.18 & 0.48$\pm$0.08 & 1.73$\pm$0.21 &  &  &  & \tabularnewline
Medium ECS GCMs & 0.77$\pm$0.10 & 1.07$\pm$0.15 & 0.66$\pm$0.08 & 1.63$\pm$0.13 &  &  &  & \tabularnewline
High ECS GCMs & 0.92$\pm$0.21 & 1.31$\pm$0.28 & 0.78$\pm$0.19 & 1.71$\pm$0.22 &  &  &  & \tabularnewline
\hline 
(60${^\circ}$S:80${^\circ}$N) & Total (${^\circ}$C) & Land (${^\circ}$C) & Ocean (${^\circ}$C) & ratio & Total (${^\circ}$C) & Land (${^\circ}$C) & Ocean (${^\circ}$C) & average ratio\tabularnewline
HadCRUT5 & 0.59 & 0.91 & 0.47 & 1.95 & 0.56 & 0.82 & 0.47 & 1.74\tabularnewline
ERA5-T2m & 0.61 & 0.97 & 0.47 & 2.08 & 0.57 & 0.82 & 0.47 & 1.74\tabularnewline
GISTEMP & 0.55 & 0.94 & 0.42 & 2.24 & 0.50 & 0.73 & 0.42 & 1.74\tabularnewline
NOAAGT & 0.54 & 0.91 & 0.39 & 2.32 & 0.47 & 0.68 & 0.39 & 1.74\tabularnewline
MSU-UAH & 0.42 & 0.55 & 0.37 & 1.51 & 0.42 & 0.54 & 0.37 & 1.74\tabularnewline
Low ECS GCMs & 0.59$\pm$0.11 & 0.86$\pm$0.19 & 0.49$\pm$0.08 & 1.75$\pm$0.20 &  &  &  & \tabularnewline
Medium ECS GCMs & 0.77$\pm$0.11 & 1.10$\pm$0.16 & 0.65$\pm$0.09 & 1.69$\pm$0.14 &  &  &  & \tabularnewline
High ECS GCMs & 0.92$\pm$0.21 & 1.34$\pm$0.29 & 0.76$\pm$0.19 & 1.79$\pm$0.24 &  &  &  & \tabularnewline
\hline 
(0${^\circ}$N:80${^\circ}$N) & Total (${^\circ}$C) & Land (${^\circ}$C) & Ocean (${^\circ}$C) & ratio & Total (${^\circ}$C) & Land (${^\circ}$C) & Ocean (${^\circ}$C) & average ratio\tabularnewline
HadCRUT5 & 0.81 & 1.03 & 0.68 & 1.52 & 0.79 & 0.99 & 0.68 & 1.45\tabularnewline
ERA5-T2m & 0.85 & 1.09 & 0.69 & 1.57 & 0.82 & 1.00 & 0.69 & 1.45\tabularnewline
GISTEMP & 0.78 & 1.04 & 0.63 & 1.66 & 0.73 & 0.91 & 0.63 & 1.45\tabularnewline
NOAAGT & 0.75 & 1.00 & 0.59 & 1.72 & 0.69 & 0.86 & 0.59 & 1.45\tabularnewline
MSU-UAH & 0.48 & 0.56 & 0.42 & 1.32 & 0.50 & 0.61 & 0.42 & 1.45\tabularnewline
Low ECS GCMs & 0.77$\pm$0.18 & 0.95$\pm$0.23 & 0.66$\pm$0.16 & 1.45$\pm$0.12 &  &  &  & \tabularnewline
Medium ECS GCMs & 1.00$\pm$0.17 & 1.21$\pm$0.19 & 0.87$\pm$0.16 & 1.40$\pm$0.09 &  &  &  & \tabularnewline
High ECS GCMs & 1.20$\pm$0.32 & 1.47$\pm$0.35 & 1.02$\pm$0.31 & 1.49$\pm$0.28 &  &  &  & \tabularnewline
\hline 
(60${^\circ}$S:0${^\circ}$S) & Total (${^\circ}$C) & Land (${^\circ}$C) & Ocean (${^\circ}$C) & ratio & Total (${^\circ}$C) & Land (${^\circ}$C) & Ocean (${^\circ}$C) & average ratio\tabularnewline
HadCRUT5 & 0.34 & 0.58 & 0.29 & 1.97 & 0.33 & 0.50 & 0.29 & 1.72\tabularnewline
ERA5-T2m & 0.34 & 0.64 & 0.29 & 2.25 & 0.32 & 0.50 & 0.29 & 1.72\tabularnewline
GISTEMP & 0.30 & 0.61 & 0.26 & 2.40 & 0.28 & 0.45 & 0.26 & 1.72\tabularnewline
NOAAGT & 0.31 & 0.65 & 0.24 & 2.68 & 0.27 & 0.41 & 0.24 & 1.72\tabularnewline
MSU-UAH & 0.35 & 0.54 & 0.32 & 1.67 & 0.34 & 0.46 & 0.32 & 1.72\tabularnewline
Low ECS GCMs & 0.39$\pm$0.08 & 0.60$\pm$0.13 & 0.35$\pm$0.08 & 1.75$\pm$0.41 &  &  &  & \tabularnewline
Medium ECS GCMs & 0.51$\pm$0.08 & 0.76$\pm$0.12 & 0.47$\pm$0.08 & 1.64$\pm$0.29 &  &  &  & \tabularnewline
High ECS GCMs & 0.61$\pm$0.13 & 0.96$\pm$0.33 & 0.55$\pm$0.11 & 1.77$\pm$0.54 &  &  &  & \tabularnewline
\hline 
\end{tabular}\caption{Left columns: observed and hindcasted warming over 80${^\circ}$S:80${^\circ}$N,
60${^\circ}$S:80${^\circ}$N, 0${^\circ}$N:80${^\circ}$N, and 60${^\circ}$S:0${^\circ}$S
latitude ranges from 1980-1990 to 2011-2021 over land+ocean (total),
land, and ocean, and land/ocean ratio. The model estimates use the
average GCM simulations. Right columns: global and land warming calculated
assuming correct the ocean warming reported by the temperature records
and the average land/ocean warming ratios hindcasted by the GCMs.}
 \label{tab4}
\end{table}

\subsection{Statistical modeling of the GCM unforced internal variability}

Figure \ref{fig5} shows the boxplots relating to the simulations
shown in Figure \ref{fig4} for each model. Again, the GCM group with
low ECS is best centered around the surface-based observations indicated
by the horizontal blue lines while the GCM groups with medium and
high ECS exhibit systematic warming bias except for very few models.
However, the dispersion of the boxplots varies greatly among the GCMs
because the models are not physically equivalent to each other and,
furthermore, probably because of the different number of simulations
available for each model.

In fact, the GCMs are represented unevenly in the KNMI collection
because the number of simulations available for each GCM varies from
3 to 100 among the models: see Figure \ref{fig5}. Therefore, the
statistics discussed in Section 3.2 may be skewed towards models with
a larger number of available simulations  because they will weight
more in the statistical test reported in Table \ref{tab3}. This problem
could be solved by using a Monte Carlo strategy to simulate the spread
of GCM hindcasts that could be associated with unforced internal 
variability. This exercise is proposed below.

It can be assumed that each GCM produces simulations distributed around
a mean $\mu_{m}$ with a given standard deviation $\sigma_{m}$ characterizing
its internal variability. We note that $\sigma_{m}$ should be assumed
constant for all GCM averages because it could be interpreted as a
``precision'' requirement for GCMs. Indeed, GCM hindcasts should
always agree with observations within an acceptable statistical uncertainty.

We propose three different options for $\sigma_{m}$ covering approximately
the ranges of the GCM boxplots shown in Figure \ref{fig5}: $\sigma_{H}\approx0.05$${^\circ}$C
(high precision), $\sigma_{M}\approx0.10$${^\circ}$C (medium precision),
and $\sigma_{L}\approx0.15$${^\circ}$C (low precision).

Figure \ref{fig5} suggests that the high-precision option ($\sigma_{H}\approx0.05$${^\circ}$C)
could be satisfied by most GCMs; it requires the model mean to be
within $\pm0.1$${^\circ}$C (95\% confidence interval) of the actual
warming value. The 95\% confidence range becomes $\pm0.2$${^\circ}$C
for the medium-precision requirement ($\sigma_{M}\approx0.10$${^\circ}$C)
and $\pm0.3$${^\circ}$C for the low-precision option ($\sigma_{L}\approx0.15$${^\circ}$C).

Appendix B shows that the interval $\pm0.1$${^\circ}$C (95\% confidence),
which corresponds to the high precision option, $\sigma_{H}\approx0.05$${^\circ}$C,
should be the preferred choice for the acceptable uncertainty related
to the internal variability that should be requested for the GCM 
because it could be derived from the variability of the temperature
records themselves.

Figure \ref{fig5} also shows that the low-precision option $\sigma_{L}\approx0.15$${^\circ}$C
is only consistent with the EC-Earth3 GCM. The usefulness of this
model should be questioned because it hindcasts 2011-2021 global surface
warming values ranging between 0.5${^\circ}$C to 1.2${^\circ}$C
with an average of 0.82${^\circ}$C. This means that EC-Earth3 is
both inaccurate and imprecise in hindcasting the global surface warming
from 1980 to 2021.

Figure \ref{fig6} shows the combined probability density functions
(PDF) and the related boxplots derived from all the GCM means reported
in Figure \ref{fig3} and in Table \ref{Tab1} with the three precision
requirements for the three ECS GCM groups compared to the warming
levels obtained with the adopted five temperature records. The complementary
Gaussian error function was used to evaluate the relative statistical
position of the five actual warming values within each probability
density function.

For each model mean $\mu_{m}$ and precision $\sigma$, the probability
$P_{m}$ that the GCM hindcast is larger than the measured warming
$\Delta T$ is

\begin{equation}
P_{m}=\frac{1}{\sigma\sqrt{2\pi}}\intop_{\Delta T}^{\infty}e^{-\frac{(t-\mu_{m})}{2\sigma^{2}}^{2}}\,dt=\frac{1}{2}\,\mathrm{erfc}\left(\frac{\Delta T-\mu_{m}}{\sigma\sqrt{2}}\right).
\end{equation}
Thus, the mean $P_{\Delta T<GCMs}=\frac{1}{N}\sum_{m=1}^{N}P_{m}$
across all models for each ECS GCM group gives the probability of
obtaining simulations warmer than the reference temperature value.
$P_{\Delta T<GCMs}$ can also be obtained by integrating the probability
density functions shown in Figure \ref{fig6}a-6c from the green line
to infinity or by using a Monte Carlo strategy by generating, for
example, 1000 computer values from a Gaussian distribution with mean
$\mu_{m}$ and standard deviation $\sigma$. The relevant statistics
are shown in Table \ref{tab3}.

Figure \ref{fig6}a-6c show that the GCM group with low ECS (blue
curves) always produces predictions well-centered on the observed
warming for the four surface temperature records because their 2011-2021
values always fall within the $\pm1\sigma$ statistical interval (which
corresponds to the 16\%-84\% probability range) of the GCM distributions
for the high, medium, and low precision options, respectively. However,
once again, if the actual 1980-2021 warming is given by UAH-MSU-lt,
even the GCM group with low ECS seems to be biased towards too hot
values in 95\%, 91\% and 85\% of possible cases, respectively, for
the three precision options (Table \ref{tab3}).

The predictions of the medium (purple) and high (red) ECS GCM groups
always show significant warming biases. Also, particularly for the
GCM group with high ECS, the PDF appears to have two peaks, implying
that the GCMs in this group are physically very different from each
other because they produce very different warming hindcasts that are
clustered around 0.8${^\circ}$C and 1.2${^\circ}$C; the warmest
PDF peak is mostly due to the CanESM5 GCM.

For the high-precision requirement ($\sigma_{H}=0.05$${^\circ}$C),
these two GCM groups produce results warmer than the observed values
from a minimum of 98\% to a maximum of 100\% of their possible output,
which is outside the 95\% confidence interval. For the medium precision
option ($\sigma_{M}=0.10$${^\circ}$C), the medium and high GCM groups
produce results warmer than the observed values from a minimum of
93\% to a maximum of 100\% of their possible outputs, which is at
the limit of the 95\% confidence interval. For the low precision option
($\sigma_{L}=0.15$${^\circ}$C), the GCM groups with medium and high
ECS produce warmer results than the four surface-based temperature
records from 88\% to 95\% of cases. Conversely, 99\% or more of the
theoretical hindcasts of the GCM groups with high and medium ECS would
be warmer than UAH-MSU-lt even for the low precision option ($\sigma_{L}=0.15$${^\circ}$C).

The boxplots illustrated in Figures \ref{fig6}d-6f were obtained
using the Monte Carlo strategy proposed above which simulates 1000
randomly distributed outputs for each of the 143 model averages for
each of the three precision options (for a total of $3\times143,000$
theoretical hindcasts). The three panels show that in all cases, with
respect to the observed temperature values, the groups with medium
and high ECS are well outside the 68\% confidence interval (i.e. the
$\pm1\sigma$ interval). Furthermore, the GCM groups with medium and
high ECS indicate levels of warming that are respectively 30\% and
50\% greater than those actually observed and, consequently, their
accuracy is rather low. The accuracy of the low-ECS GCM group is good
compared to the surface-based temperature records, but it still reports
average warming that is about 30\% larger than that reported by the
satellite temperature record. The whisker extension of the boxplot
shows that the precision of the low, medium and high ECS groups varies
from modest ($\pm0.2$${^\circ}$C) to very poor ($\pm0.5$${^\circ}$C)
range from low ECS and high precision GCM group to high and low precision
GCM group.

\begin{figure}[!t]
\centering{}\includegraphics[width=1\textwidth]{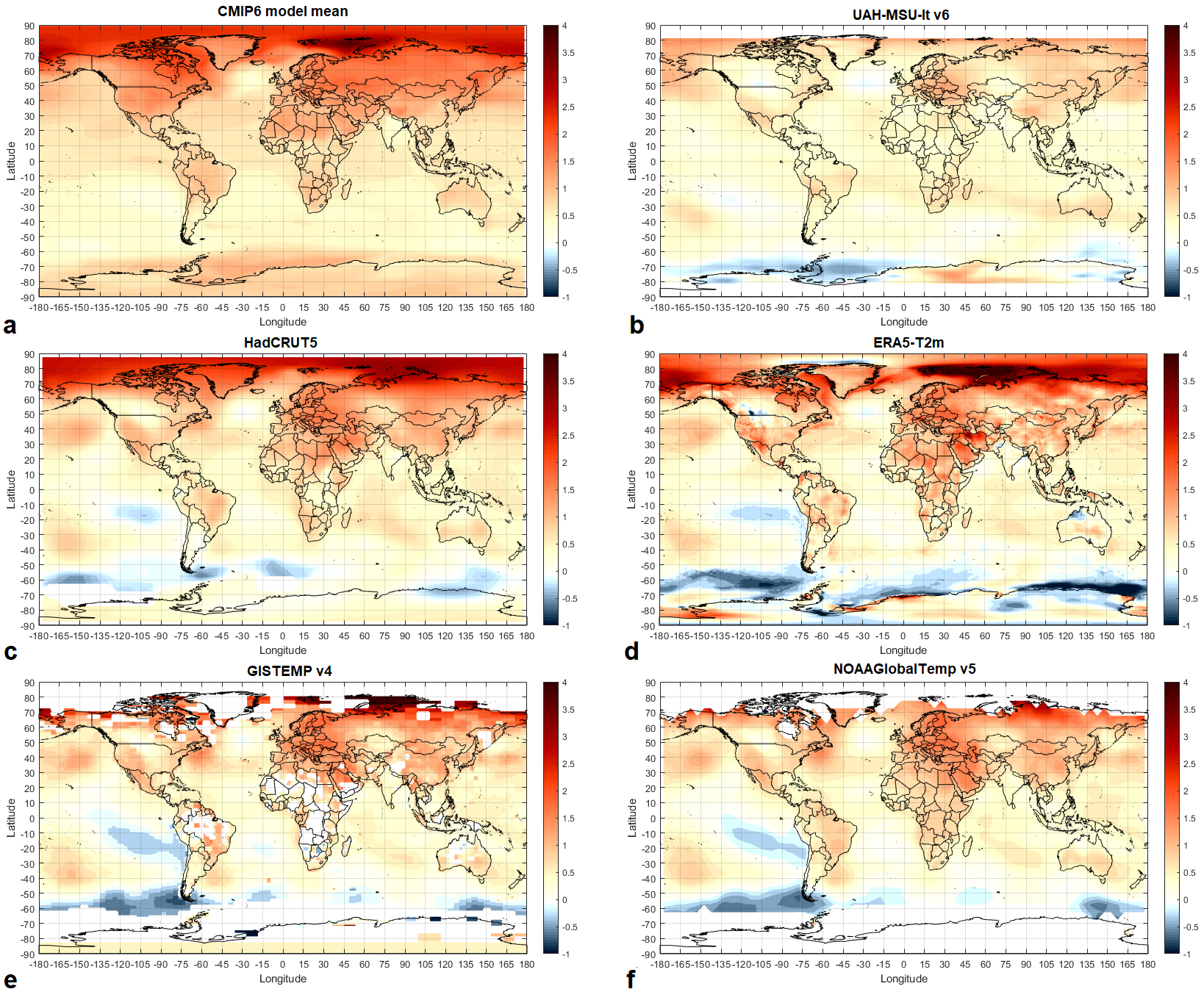}\caption{Areal distribution of the warming from 1980-1990 to 2011-2021 for
the CMIP6 ensemble average simulation and for the HadCRUT5, ERA5-T2m,
GISTEMP v4, NOAAGlobTemp v5, and UAH-MSU-lt v6 temperature records.}
\label{fig7}
\end{figure}

\begin{figure}[!t]
\centering{}\includegraphics[width=1\textwidth]{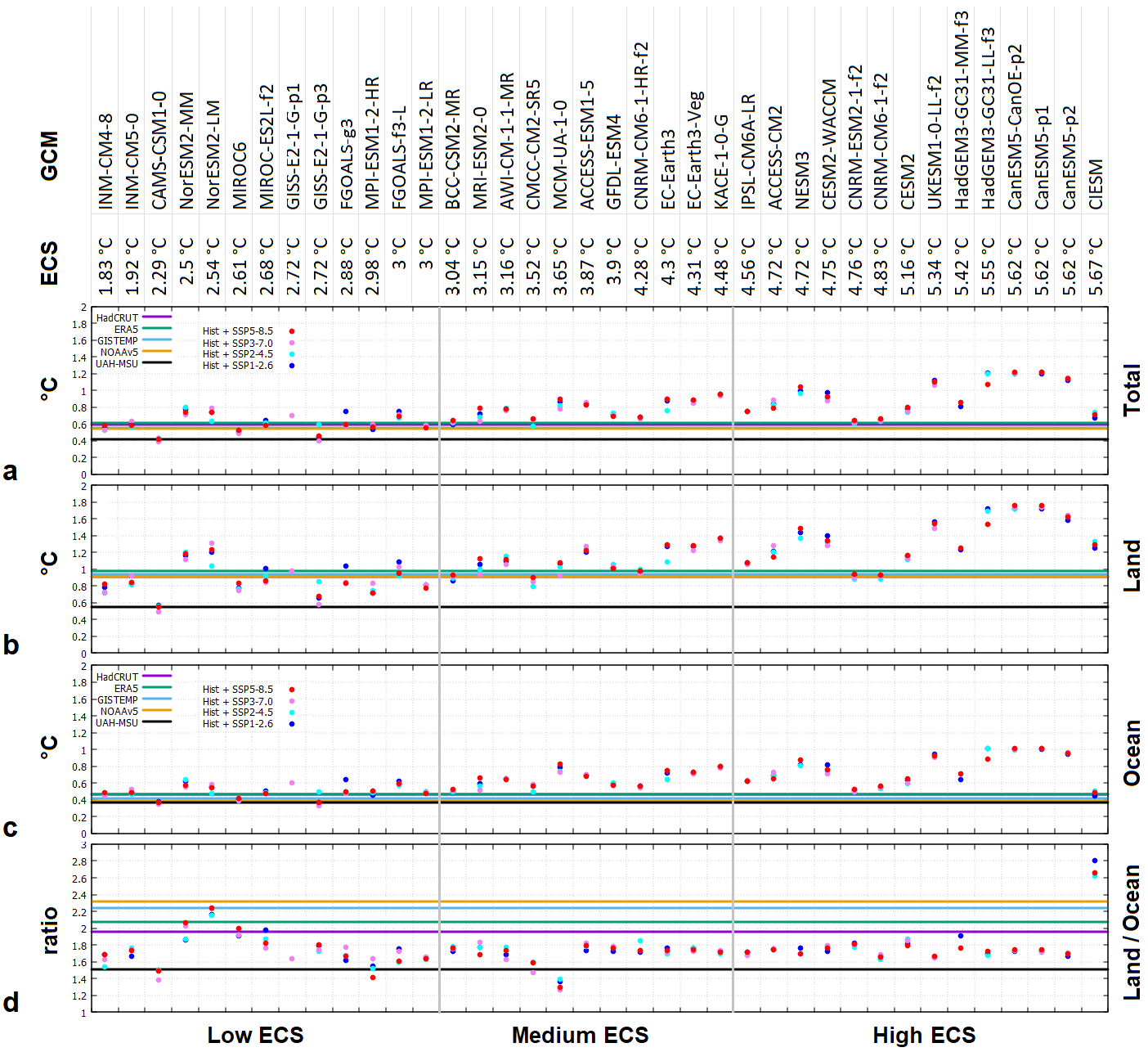}\caption{2011-2021 warming relative to 1980-1990 of the models over (a) land+ocean,
(b) land, (c) ocean areas within the 60${^\circ}$S:80${^\circ}$N
latitude range for the 143 average model simulations (colored dots)
and the five temperature records (colored lines). (d) Ratio between
the land and the ocean warming. See Tables \ref{tab5}, \ref{tab6},
and \ref{tab7}.}
\label{fig8}
\end{figure}

\begin{table}
\centering{}%
\begin{tabular}{lcccccc}
﻿Model & ECS (${^\circ}$C) & Hist + SSP & Land + Ocean (${^\circ}$C) & Land (${^\circ}$C) & Ocean (${^\circ}$C) & Land/Ocean\tabularnewline
\hline 
INM-CM4-8 & 1.83 & SSP1-2.6 & 0.55 & 0.78 & 0.46 & 1.69\tabularnewline
INM-CM4-8 & 1.83 & SSP2-4.5 & 0.54 & 0.72 & 0.47 & 1.54\tabularnewline
INM-CM4-8 & 1.83 & SSP3-7.0 & 0.52 & 0.72 & 0.45 & 1.62\tabularnewline
INM-CM4-8 & 1.83 & SSP5-8.5 & 0.58 & 0.82 & 0.49 & 1.68\tabularnewline
INM-CM5-0 & 1.92 & SSP1-2.6 & 0.59 & 0.83 & 0.50 & 1.66\tabularnewline
INM-CM5-0 & 1.92 & SSP2-4.5 & 0.56 & 0.82 & 0.46 & 1.76\tabularnewline
INM-CM5-0 & 1.92 & SSP3-7.0 & 0.64 & 0.92 & 0.53 & 1.74\tabularnewline
INM-CM5-0 & 1.92 & SSP5-8.5 & 0.59 & 0.84 & 0.49 & 1.73\tabularnewline
CAMS-CSM1-0 & 2.29 & SSP1-2.6 & 0.43 & 0.57 & 0.38 & 1.50\tabularnewline
CAMS-CSM1-0 & 2.29 & SSP2-4.5 & 0.42 & 0.56 & 0.37 & 1.50\tabularnewline
CAMS-CSM1-0 & 2.29 & SSP3-7.0 & 0.39 & 0.49 & 0.35 & 1.39\tabularnewline
CAMS-CSM1-0 & 2.29 & SSP5-8.5 & 0.42 & 0.55 & 0.37 & 1.49\tabularnewline
NorESM2-MM & 2.5 & SSP1-2.6 & 0.77 & 1.16 & 0.62 & 1.86\tabularnewline
NorESM2-MM & 2.5 & SSP2-4.5 & 0.80 & 1.20 & 0.64 & 1.87\tabularnewline
NorESM2-MM & 2.5 & SSP3-7.0 & 0.71 & 1.11 & 0.55 & 2.03\tabularnewline
NorESM2-MM & 2.5 & SSP5-8.5 & 0.74 & 1.18 & 0.57 & 2.06\tabularnewline
NorESM2-LM & 2.54 & SSP1-2.6 & 0.74 & 1.21 & 0.56 & 2.17\tabularnewline
NorESM2-LM & 2.54 & SSP2-4.5 & 0.64 & 1.04 & 0.48 & 2.15\tabularnewline
NorESM2-LM & 2.54 & SSP3-7.0 & 0.79 & 1.31 & 0.59 & 2.23\tabularnewline
NorESM2-LM & 2.54 & SSP5-8.5 & 0.74 & 1.23 & 0.55 & 2.24\tabularnewline
MIROC6 & 2.61 & SSP1-2.6 & 0.51 & 0.78 & 0.41 & 1.91\tabularnewline
MIROC6 & 2.61 & SSP2-4.5 & 0.50 & 0.77 & 0.40 & 1.92\tabularnewline
MIROC6 & 2.61 & SSP3-7.0 & 0.49 & 0.75 & 0.39 & 1.92\tabularnewline
MIROC6 & 2.61 & SSP5-8.5 & 0.53 & 0.83 & 0.41 & 2.00\tabularnewline
MIROC-ES2L-f2 & 2.68 & SSP1-2.6 & 0.65 & 1.00 & 0.51 & 1.98\tabularnewline
MIROC-ES2L-f2 & 2.68 & SSP2-4.5 & 0.61 & 0.92 & 0.49 & 1.87\tabularnewline
MIROC-ES2L-f2 & 2.68 & SSP3-7.0 & 0.58 & 0.84 & 0.48 & 1.76\tabularnewline
MIROC-ES2L-f2 & 2.68 & SSP5-8.5 & 0.58 & 0.87 & 0.48 & 1.82\tabularnewline
GISS-E2-1-G-p1 & 2.72 & SSP3-7.0 & 0.70 & 0.98 & 0.60 & 1.63\tabularnewline
GISS-E2-1-G-p3 & 2.72 & SSP1-2.6 & 0.45 & 0.66 & 0.37 & 1.79\tabularnewline
GISS-E2-1-G-p3 & 2.72 & SSP2-4.5 & 0.60 & 0.86 & 0.50 & 1.72\tabularnewline
GISS-E2-1-G-p3 & 2.72 & SSP3-7.0 & 0.40 & 0.58 & 0.33 & 1.74\tabularnewline
GISS-E2-1-G-p3 & 2.72 & SSP5-8.5 & 0.46 & 0.67 & 0.37 & 1.80\tabularnewline
FGOALS-g3 & 2.88 & SSP1-2.6 & 0.75 & 1.04 & 0.64 & 1.62\tabularnewline
FGOALS-g3 & 2.88 & SSP2-4.5 & 0.59 & 0.84 & 0.50 & 1.67\tabularnewline
FGOALS-g3 & 2.88 & SSP3-7.0 & 0.58 & 0.84 & 0.47 & 1.77\tabularnewline
FGOALS-g3 & 2.88 & SSP5-8.5 & 0.59 & 0.84 & 0.50 & 1.67\tabularnewline
MPI-ESM1-2-HR & 2.98 & SSP1-2.6 & 0.53 & 0.71 & 0.46 & 1.55\tabularnewline
MPI-ESM1-2-HR & 2.98 & SSP2-4.5 & 0.56 & 0.75 & 0.49 & 1.52\tabularnewline
MPI-ESM1-2-HR & 2.98 & SSP3-7.0 & 0.60 & 0.83 & 0.51 & 1.64\tabularnewline
MPI-ESM1-2-HR & 2.98 & SSP5-8.5 & 0.56 & 0.71 & 0.50 & 1.41\tabularnewline
FGOALS-f3-L & 3 & SSP1-2.6 & 0.75 & 1.08 & 0.62 & 1.75\tabularnewline
FGOALS-f3-L & 3 & SSP2-4.5 & 0.67 & 0.92 & 0.57 & 1.60\tabularnewline
FGOALS-f3-L & 3 & SSP3-7.0 & 0.71 & 1.02 & 0.59 & 1.72\tabularnewline
FGOALS-f3-L & 3 & SSP5-8.5 & 0.69 & 0.95 & 0.59 & 1.61\tabularnewline
MPI-ESM1-2-LR & 3 & SSP1-2.6 & 0.58 & 0.81 & 0.49 & 1.65\tabularnewline
MPI-ESM1-2-LR & 3 & SSP2-4.5 & 0.58 & 0.81 & 0.50 & 1.63\tabularnewline
MPI-ESM1-2-LR & 3 & SSP3-7.0 & 0.58 & 0.81 & 0.49 & 1.66\tabularnewline
MPI-ESM1-2-LR & 3 & SSP5-8.5 & 0.56 & 0.78 & 0.47 & 1.64\tabularnewline
mean &  &  & 0.59 $\pm$ 0.11 & 0.86$\pm$ 0.19 & 0.49$\pm$ 0.08 & 1.75$\pm$ 0.20\tabularnewline
\end{tabular}\caption{Low-ECS GCMs: hindcasted warming from 1980-1990 to 2011-2021 within
the 60${^\circ}$S:80${^\circ}$N latitude range from 1980-1990 to
2011-2021 over land+ocean (total), land, ocean, and land/ocean ratio.
See Figure \ref{fig8}.}
 \label{tab5}
\end{table}

\begin{table}
\centering{}%
\begin{tabular}{lcccccc}
﻿Model & ECS (${^\circ}$C) & Hist + SSP & Land + Ocean (${^\circ}$C) & Land (${^\circ}$C) & Ocean (${^\circ}$C) & Land/Ocean\tabularnewline
\hline 
BCC-CSM2-MR & 3.04 & SSP1-2.6 & 0.60 & 0.86 & 0.50 & 1.73\tabularnewline
BCC-CSM2-MR & 3.04 & SSP2-4.5 & 0.61 & 0.89 & 0.50 & 1.79\tabularnewline
BCC-CSM2-MR & 3.04 & SSP3-7.0 & 0.63 & 0.91 & 0.52 & 1.75\tabularnewline
BCC-CSM2-MR & 3.04 & SSP5-8.5 & 0.64 & 0.93 & 0.53 & 1.76\tabularnewline
MRI-ESM2-0 & 3.15 & SSP1-2.6 & 0.72 & 1.05 & 0.59 & 1.77\tabularnewline
MRI-ESM2-0 & 3.15 & SSP2-4.5 & 0.68 & 1.00 & 0.56 & 1.77\tabularnewline
MRI-ESM2-0 & 3.15 & SSP3-7.0 & 0.63 & 0.94 & 0.51 & 1.83\tabularnewline
MRI-ESM2-0 & 3.15 & SSP5-8.5 & 0.79 & 1.12 & 0.66 & 1.69\tabularnewline
AWI-CM-1-1-MR & 3.16 & SSP1-2.6 & 0.77 & 1.09 & 0.65 & 1.69\tabularnewline
AWI-CM-1-1-MR & 3.16 & SSP2-4.5 & 0.79 & 1.15 & 0.65 & 1.78\tabularnewline
AWI-CM-1-1-MR & 3.16 & SSP3-7.0 & 0.76 & 1.05 & 0.65 & 1.62\tabularnewline
AWI-CM-1-1-MR & 3.16 & SSP5-8.5 & 0.77 & 1.11 & 0.64 & 1.73\tabularnewline
CMCC-CM2-SR5 & 3.52 & SSP2-4.5 & 0.58 & 0.79 & 0.50 & 1.60\tabularnewline
CMCC-CM2-SR5 & 3.52 & SSP3-7.0 & 0.66 & 0.85 & 0.58 & 1.47\tabularnewline
CMCC-CM2-SR5 & 3.52 & SSP5-8.5 & 0.66 & 0.90 & 0.57 & 1.59\tabularnewline
MCM-UA-1-0 & 3.65 & SSP1-2.6 & 0.86 & 1.07 & 0.78 & 1.36\tabularnewline
MCM-UA-1-0 & 3.65 & SSP2-4.5 & 0.82 & 1.03 & 0.74 & 1.40\tabularnewline
MCM-UA-1-0 & 3.65 & SSP3-7.0 & 0.78 & 0.92 & 0.73 & 1.27\tabularnewline
MCM-UA-1-0 & 3.65 & SSP5-8.5 & 0.90 & 1.07 & 0.83 & 1.29\tabularnewline
ACCESS-ESM1-5 & 3.87 & SSP1-2.6 & 0.84 & 1.20 & 0.69 & 1.74\tabularnewline
ACCESS-ESM1-5 & 3.87 & SSP2-4.5 & 0.85 & 1.24 & 0.70 & 1.78\tabularnewline
ACCESS-ESM1-5 & 3.87 & SSP3-7.0 & 0.86 & 1.27 & 0.70 & 1.82\tabularnewline
ACCESS-ESM1-5 & 3.87 & SSP5-8.5 & 0.83 & 1.22 & 0.68 & 1.80\tabularnewline
GFDL-ESM4 & 3.9 & SSP1-2.6 & 0.71 & 1.01 & 0.59 & 1.72\tabularnewline
GFDL-ESM4 & 3.9 & SSP2-4.5 & 0.73 & 1.06 & 0.60 & 1.77\tabularnewline
GFDL-ESM4 & 3.9 & SSP3-7.0 & 0.70 & 1.02 & 0.57 & 1.78\tabularnewline
GFDL-ESM4 & 3.9 & SSP5-8.5 & 0.69 & 1.00 & 0.57 & 1.76\tabularnewline
CNRM-CM6-1-HR-f2 & 4.28 & SSP1-2.6 & 0.68 & 0.97 & 0.57 & 1.71\tabularnewline
CNRM-CM6-1-HR-f2 & 4.28 & SSP2-4.5 & 0.67 & 1.00 & 0.54 & 1.85\tabularnewline
CNRM-CM6-1-HR-f2 & 4.28 & SSP3-7.0 & 0.66 & 0.95 & 0.55 & 1.73\tabularnewline
CNRM-CM6-1-HR-f2 & 4.28 & SSP5-8.5 & 0.68 & 0.98 & 0.57 & 1.73\tabularnewline
EC-Earth3 & 4.3 & SSP1-2.6 & 0.87 & 1.27 & 0.72 & 1.76\tabularnewline
EC-Earth3 & 4.3 & SSP2-4.5 & 0.76 & 1.08 & 0.64 & 1.69\tabularnewline
EC-Earth3 & 4.3 & SSP3-7.0 & 0.90 & 1.29 & 0.75 & 1.71\tabularnewline
EC-Earth3 & 4.3 & SSP5-8.5 & 0.90 & 1.29 & 0.75 & 1.73\tabularnewline
EC-Earth3-Veg & 4.31 & SSP1-2.6 & 0.87 & 1.27 & 0.72 & 1.76\tabularnewline
EC-Earth3-Veg & 4.31 & SSP2-4.5 & 0.88 & 1.28 & 0.72 & 1.76\tabularnewline
EC-Earth3-Veg & 4.31 & SSP3-7.0 & 0.85 & 1.22 & 0.71 & 1.73\tabularnewline
EC-Earth3-Veg & 4.31 & SSP5-8.5 & 0.88 & 1.28 & 0.73 & 1.75\tabularnewline
KACE-1-0-G & 4.48 & SSP1-2.6 & 0.95 & 1.37 & 0.79 & 1.73\tabularnewline
KACE-1-0-G & 4.48 & SSP2-4.5 & 0.96 & 1.36 & 0.80 & 1.69\tabularnewline
KACE-1-0-G & 4.48 & SSP3-7.0 & 0.93 & 1.34 & 0.77 & 1.73\tabularnewline
KACE-1-0-G & 4.48 & SSP5-8.5 & 0.96 & 1.37 & 0.80 & 1.72\tabularnewline
mean &  &  & 0.77 $\pm$ 0.11 & 1.10 $\pm$ 0.16 & 0.65 $\pm$ 0.10 & 1.69 $\pm$ 0.14\tabularnewline
\end{tabular}\caption{Medium-ECS GCMs: hindcasted warming from 1980-1990 to 2011-2021 within
the 60${^\circ}$S:80${^\circ}$N latitude range from 1980-1990 to
2011-2021 over land+ocean (total), land, ocean, and land/ocean ratio.
See Figure \ref{fig8}.}
 \label{tab6}
\end{table}

\begin{table}
\centering{}%
\begin{tabular}{lcccccc}
﻿Model & ECS (${^\circ}$C) & Hist + SSP & Land + Ocean (${^\circ}$C) & Land (${^\circ}$C) & Ocean (${^\circ}$C) & Land/Ocean\tabularnewline
\hline 
IPSL-CM6A-LR & 4.56 & SSP1-2.6 & 0.75 & 1.07 & 0.63 & 1.71\tabularnewline
IPSL-CM6A-LR & 4.56 & SSP2-4.5 & 0.75 & 1.07 & 0.63 & 1.70\tabularnewline
IPSL-CM6A-LR & 4.56 & SSP3-7.0 & 0.75 & 1.06 & 0.63 & 1.68\tabularnewline
IPSL-CM6A-LR & 4.56 & SSP5-8.5 & 0.75 & 1.08 & 0.63 & 1.72\tabularnewline
ACCESS-CM2 & 4.72 & SSP1-2.6 & 0.84 & 1.21 & 0.69 & 1.75\tabularnewline
ACCESS-CM2 & 4.72 & SSP2-4.5 & 0.83 & 1.20 & 0.69 & 1.74\tabularnewline
ACCESS-CM2 & 4.72 & SSP3-7.0 & 0.88 & 1.28 & 0.73 & 1.76\tabularnewline
ACCESS-CM2 & 4.72 & SSP5-8.5 & 0.79 & 1.14 & 0.65 & 1.74\tabularnewline
NESM3 & 4.72 & SSP1-2.6 & 0.99 & 1.44 & 0.82 & 1.76\tabularnewline
NESM3 & 4.72 & SSP2-4.5 & 0.97 & 1.37 & 0.81 & 1.69\tabularnewline
NESM3 & 4.72 & SSP5-8.5 & 1.04 & 1.48 & 0.87 & 1.70\tabularnewline
CESM2-WACCM & 4.75 & SSP1-2.6 & 0.98 & 1.40 & 0.81 & 1.72\tabularnewline
CESM2-WACCM & 4.75 & SSP2-4.5 & 0.90 & 1.32 & 0.74 & 1.78\tabularnewline
CESM2-WACCM & 4.75 & SSP3-7.0 & 0.87 & 1.28 & 0.71 & 1.80\tabularnewline
CESM2-WACCM & 4.75 & SSP5-8.5 & 0.92 & 1.34 & 0.76 & 1.77\tabularnewline
CNRM-ESM2-1-f2 & 4.76 & SSP1-2.6 & 0.63 & 0.93 & 0.51 & 1.82\tabularnewline
CNRM-ESM2-1-f2 & 4.76 & SSP2-4.5 & 0.60 & 0.88 & 0.50 & 1.77\tabularnewline
CNRM-ESM2-1-f2 & 4.76 & SSP3-7.0 & 0.61 & 0.90 & 0.50 & 1.81\tabularnewline
CNRM-ESM2-1-f2 & 4.76 & SSP5-8.5 & 0.64 & 0.94 & 0.52 & 1.81\tabularnewline
CNRM-CM6-1-f2 & 4.83 & SSP1-2.6 & 0.66 & 0.92 & 0.55 & 1.67\tabularnewline
CNRM-CM6-1-f2 & 4.83 & SSP2-4.5 & 0.63 & 0.88 & 0.54 & 1.62\tabularnewline
CNRM-CM6-1-f2 & 4.83 & SSP3-7.0 & 0.65 & 0.91 & 0.54 & 1.68\tabularnewline
CNRM-CM6-1-f2 & 4.83 & SSP5-8.5 & 0.66 & 0.93 & 0.56 & 1.66\tabularnewline
CESM2 & 5.16 & SSP1-2.6 & 0.78 & 1.15 & 0.64 & 1.82\tabularnewline
CESM2 & 5.16 & SSP2-4.5 & 0.74 & 1.11 & 0.60 & 1.87\tabularnewline
CESM2 & 5.16 & SSP3-7.0 & 0.75 & 1.13 & 0.61 & 1.86\tabularnewline
CESM2 & 5.16 & SSP5-8.5 & 0.79 & 1.17 & 0.65 & 1.79\tabularnewline
UKESM1-0-LL-f2 & 5.34 & SSP1-2.6 & 1.12 & 1.56 & 0.94 & 1.66\tabularnewline
UKESM1-0-LL-f2 & 5.34 & SSP2-4.5 & 1.10 & 1.53 & 0.93 & 1.65\tabularnewline
UKESM1-0-LL-f2 & 5.34 & SSP3-7.0 & 1.06 & 1.48 & 0.90 & 1.64\tabularnewline
UKESM1-0-LL-f2 & 5.34 & SSP5-8.5 & 1.10 & 1.55 & 0.93 & 1.67\tabularnewline
HadGEM3-GC31-MM-f3 & 5.42 & SSP1-2.6 & 0.81 & 1.23 & 0.64 & 1.91\tabularnewline
HadGEM3-GC31-MM-f3 & 5.42 & SSP5-8.5 & 0.86 & 1.25 & 0.71 & 1.76\tabularnewline
HadGEM3-GC31-LL-f3 & 5.55 & SSP1-2.6 & 1.21 & 1.71 & 1.01 & 1.70\tabularnewline
HadGEM3-GC31-LL-f3 & 5.55 & SSP2-4.5 & 1.20 & 1.69 & 1.01 & 1.68\tabularnewline
HadGEM3-GC31-LL-f3 & 5.55 & SSP5-8.5 & 1.07 & 1.53 & 0.89 & 1.72\tabularnewline
CanESM5-CanOE-p2 & 5.62 & SSP1-2.6 & 1.20 & 1.72 & 1.00 & 1.72\tabularnewline
CanESM5-CanOE-p2 & 5.62 & SSP2-4.5 & 1.19 & 1.72 & 0.99 & 1.73\tabularnewline
CanESM5-CanOE-p2 & 5.62 & SSP3-7.0 & 1.20 & 1.73 & 1.00 & 1.74\tabularnewline
CanESM5-CanOE-p2 & 5.62 & SSP5-8.5 & 1.22 & 1.76 & 1.01 & 1.74\tabularnewline
CanESM5-p1 & 5.62 & SSP1-2.6 & 1.20 & 1.72 & 1.00 & 1.72\tabularnewline
CanESM5-p1 & 5.62 & SSP2-4.5 & 1.21 & 1.74 & 1.01 & 1.71\tabularnewline
CanESM5-p1 & 5.62 & SSP3-7.0 & 1.22 & 1.74 & 1.01 & 1.72\tabularnewline
CanESM5-p1 & 5.62 & SSP5-8.5 & 1.22 & 1.76 & 1.01 & 1.74\tabularnewline
CanESM5-p2 & 5.62 & SSP1-2.6 & 1.12 & 1.58 & 0.95 & 1.67\tabularnewline
CanESM5-p2 & 5.62 & SSP2-4.5 & 1.14 & 1.63 & 0.96 & 1.70\tabularnewline
CanESM5-p2 & 5.62 & SSP3-7.0 & 1.15 & 1.64 & 0.96 & 1.70\tabularnewline
CanESM5-p2 & 5.62 & SSP5-8.5 & 1.14 & 1.62 & 0.95 & 1.69\tabularnewline
CIESM & 5.67 & SSP1-2.6 & 0.67 & 1.25 & 0.44 & 2.81\tabularnewline
CIESM & 5.67 & SSP2-4.5 & 0.73 & 1.33 & 0.51 & 2.62\tabularnewline
CIESM & 5.67 & SSP5-8.5 & 0.71 & 1.29 & 0.48 & 2.66\tabularnewline
mean &  &  & 0.92 $\pm$ 0.21 & 1.34 $\pm$ 0.29 & 0.76 $\pm$ 0.19 & 1.79 $\pm$ 0.24\tabularnewline
\end{tabular}\caption{High-ECS GCMs: hindcasted warming from 1980-1990 to 2011-2021 within
the 60${^\circ}$S:80${^\circ}$N latitude range from 1980-1990 to
2011-2021 over land+ocean (total), land, ocean, and land/ocean ratio.
See Figure \ref{fig8}.}
 \label{tab7}
\end{table}

\section{Testing the land versus the ocean warming}

Surface-based temperature records imply that the GCM group with low
ECS performs better than those with medium and high ECS, which suggests
that the most likely ECS value should be equal or lower than 3${^\circ}$C.
However, UAH-MSU-lt implies that even the low-ECS GCMs may perform
quite poorly. The observed discrepancy between the surface and satellite
temperature records may be due to the presence of various non-climatic
warming biases in the surface temperature records \citep{Connolly,DAleo(2016),Scafetta2021,Watts(2022)}.
This problem is now being investigated by comparing the GCM hindcasts
against the land and the ocean temperature observations.

Figures \ref{fig7}a-7f show the areal distribution of warming from
1980-1990 to 2011-2021 produced by the CMIP6 GCM ensemble average
and by HadCRUT5, ERA5-T2m, GISTEMP v4, NOAAGlobTemp v5 and UAH-MSU-lt
v6. Equivalent maps for each GCM are found in \citet{ScafettaMDPI2021}.

Figure \ref{fig7}b shows that the UAH-MSU-lt v6 temperature record
covers the latitude range 80${^\circ}$S-80${^\circ}$N. Figure \ref{fig7}c
and 7d show that ERA5-T2m and HadCRUT5 (infilled data) are global
because they adopt interpolations of meteorological models to extend
coverage also in data-scattered regions of the globe such as the poles
and other inhabited areas (large deserts and forests). Figures \ref{fig7}e
and 7f show that the GISTEMP and NOAAGlobTemp records do not cover
large areas, in particular, the polar regions are poorly represented.

Figure \ref{fig7}b is characterized by lighter colors than the other
temperature panels, which means that the UAH-MSU-lt temperature record
shows less warming than the surface-based temperature records almost
everywhere. All six temperature panels in Figure \ref{fig7} also
show that the land area has warmed more than the ocean region. In
any case, Figures \ref{fig7}c-7f show that the surface temperature
records present a greater temperature difference between the land
and the ocean regions. The visual comparison with the CMIP6 ensemble
average simulation (Figure \ref{fig7}a) suggests the same general
pattern but, furthermore, the oceanic area appears slightly warmer
than all five temperature records. The temperature records also show
extensive ocean areas where significant cooling is observed such as
around Antarctica, the eastern equatorial Pacific, the North Atlantic
and a few other regions. These cooling regions reveal interesting
dynamic patterns that are not captured by the average simulation of
the CMIP6 ensemble. These patterns are best emphasized in the areal
t-test proposed in \citet{Scafetta}.

Table \ref{tab4} reports the warming over 80${^\circ}$S:80${^\circ}$N,
60${^\circ}$S:80${^\circ}$N, 0${^\circ}$N:80${^\circ}$N and 60${^\circ}$S:0${^\circ}$S
latitudinal ranges from 1980-1990 to 2011-2021 over land+ocean (total),
land, and ocean. Table \ref{tab4} also reports the ratios between
the land and the ocean warming levels.

The area 0${^\circ}$N:80${^\circ}$N shows that from 1980-1990 to
2011-2021 the surface temperature records warmed on average by about
$0.32\pm0.05$${^\circ}$C more than the satellite-based UAH-MSU-lt
record, while the area 80${^\circ}$S:80${^\circ}$N the surface-based
records warmed on average by about $0.15\pm0.02$${^\circ}$C more
than the satellite record. A similar warming bias on land also appears
in the Southern Hemisphere (60${^\circ}$S:0${^\circ}$S) because
the surface-based temperate records show ocean warming averaging $0.05\pm0.03$
${^\circ}$C less than the satellite record while their land area
warmed by $0.08\pm0.03$ ${^\circ}$C more.

Figure \ref{fig8} shows the results for each GCM model (using the
143 GCM average simulations available for each SSP) for the latitudinal
interval 60${^\circ}$S:80${^\circ}$N, which is optimally covered
from all temperatures records and includes all continents except Antarctica.
The results are also reported in Tables \ref{tab5}, \ref{tab6},
and \ref{tab7} and could be used to evaluate possible anomalous temperature
trends on the continents.

Figure \ref{fig8}a compares the synthetic and observed global warming
levels from 1980-1990 and 2011-2021. Figures \ref{fig8}b and 8c show
the land and the ocean average warming levels, respectively. These
figures show that the performance of the models is similar to what
we have obtained in the previous sections, i.e. the GCM group with
low ECS performs significantly better than the medium and high GCM
groups, which show warming bias for most of their GCMs.

Figure \ref{fig8}d shows the relationships between average warming
on the land and the ocean areas. The mean land/ocean ratio for the
vast majority of the models is $1.75\pm0.20$, which is a value placed
between the results obtained for the surface temperature records (ranging
from 1.95 to 2.32 ) and that of the satellite temperature record,
which gives 1.51.

The results shown in Figure \ref{fig8} can be interpreted as follows.

1) Figure \ref{fig8}b shows that the land surface temperature records
are on average 0.4${^\circ}$C warmer than the satellite-based one.
On the contrary, Figure \ref{fig8}c shows that the surface-based
ocean temperatures are on average up to a maximum of 0.1${^\circ}$C
warmer than the satellite ones.

2) Therefore, it can be assumed that on the ocean, the satellite-based
temperature record is sufficiently compatible with the surface-based
ones. If so, the large divergence observed on land between surface
and satellite recordings could suggest that the land measurements
are significantly contaminated by non-climatic warming biaes, including
those related to urbanization \citep[cf.:][]{Connolly,DAleo(2016),Scafetta2021,Watts(2022)}.

3) A similar conclusion would also be indirectly supported by the
GCM hindcasts which show that the CMIP6 models are usually unable
to correctly reconstruct the large land/ocean temperature ratio observed
in the surface temperature records. In fact, the models give a land/ocean
ratio equal to $1.75\pm0.20$, while the surface records give ratios
between 1.95 and 2.32.

4) However, as the GCMs attempt to reconstruct the global surface
warming of the surface temperature records even though they cannot
adequately explain their large land/ocean warming ratio, the models
could have calibrated internal parameters to obtain a compromise that
attempts to approximate the global surface warming by simulating a
warmer ocean and a cooler land than observed.

If point 4 above is correct, the reliability of the low-ECS GCMs should
also be questioned. In fact, \ref{fig8}d shows contradictory results
regarding the low-ECS GCMs because some models agree better with the
surface-based temperature records, few others agree better with the
satellite temperature record, while the rest report land/ocean ratios
between the two levels, as  the vast majority of the medium and high
ECS models does.

We now assume that the GCM's predicted land/ocean temperature ratio
(average ratio = $1.75\pm0.20$) corresponds to the actual physical
characteristics of the climate system and that the ocean temperature
warming of the surface records (on average, $0.43\pm0.03$ ${^\circ}$C,
see Table \ref{tab4}) is sufficiently accurate. If so, from 1980-1990
to 2011-2021 the earth's surface within the latitude interval 60${^\circ}$S:80${^\circ}$N
should have warmed by $0.75\pm0.1$${^\circ}$C instead of the observed
$0.93\pm0.03$${^\circ}$C. If the hypothesis is correct, the spurious
warming of the land surface due to uncorrected non-climatic warming
biases could be quantified as approximately +0.2${^\circ}$C. The
proposed correction implies that global surface warming from 1980-1990
to 2011-2021 could be at least about 0.05${^\circ}$C ($\sim$10\%)
lower than what the surface-based records report, which increases
further the warming bias of the medium and high-ECS GCMs observed
in Figures \ref{fig1}-\ref{fig6}.

The results depicted in Figure \ref{fig8} also help to better evaluate
the individual GCMs. For example, Figure \ref{fig5} suggests that
three high-ECS models (CNRM-CM6-1-f2, CNRM-ESM2-1-f2 and CIESM) produce
relatively close warming to what is reported by the surface-based
temperature records. However, Figure \ref{fig8}d indicates that the
same models fail to produce the land/ocean temperature ratio of the
same temperature records showing significantly lower (CNRM) or higher
(CIESM) results than reported. Therefore, it appears that in these
GCMs the biases that occur in some regions are offset by opposite
biases that occur in other regions.

The last four columns of Table \ref{tab4} report the global (land+ocean)
and land warming calculated assuming that the ocean warming of the
temperature records is correct and that the land/ocean warming ratios
hindcasted by the models is correct as well. The global estimate was
calculated from the ocean and the land ones weighted with their relative
area percentages within each latitudinal range. In particular, we
found that for the Northern Hemisphere (0${^\circ}$N:80${^\circ}$N),
the land could have warmed about 0.087${^\circ}$C less than what
reported on average by HadCRUT5, ERA-T2m, and GISTEMP. This bias roughly
corresponds to the different warming estimated in \citet{Connolly}
for the northern hemisphere land area by comparing the temperature
records reconstructed by using both urban+rural stations and rural-only
stations that should present significantly mitigated  non-climatic
warming biases.

In conclusion, the proposed land-ocean comparison suggests that the
surface-based temperature records most likely exhibit non-climatic
warming biases and that the actual global surface warming from 1980-1990
to 2011-2021 may have been approximately between 0.50${^\circ}$C
and 0.55${^\circ}$C, which is approximately 10\% lower than what
is reported in Section 2. This means that the medium and high-ECS
GCM groups are further confirmed to run too hot and that the low ECS
group performs slightly worse than concluded in Section 3 because
the average warming of its hindcasts from 1980-1990 to 2011-2021 is
approximately 0.6${^\circ}$C (Table \ref{Tab1}). However, if UAH-MSU-lt
reproduces the global surface warming more accurately, the surface-based
temperature records would exhibit warming bias of up to 30\% of the
reported values, which would indicate that even the low ECS GCMs run
significantly too hot and need to be scaled down by 33\% to reduce
their mean warming from 0.6${^\circ}$C to 0.4${^\circ}$C, which
is the warming reported by the satellite-based measurements.  Indeed,
another indirect evidence that the land surface temperature records
could be affected by a significant warming bias is also given by the
divergence observed between instrumental and  dendroclimatological proxy
temperature records over the past 50 years, where the former show
a warming trend significantly higher than the latter \citep{Buntgen,Esper,Scafetta2021}.

\begin{figure}[!t]
\centering{}\includegraphics[width=1\textwidth]{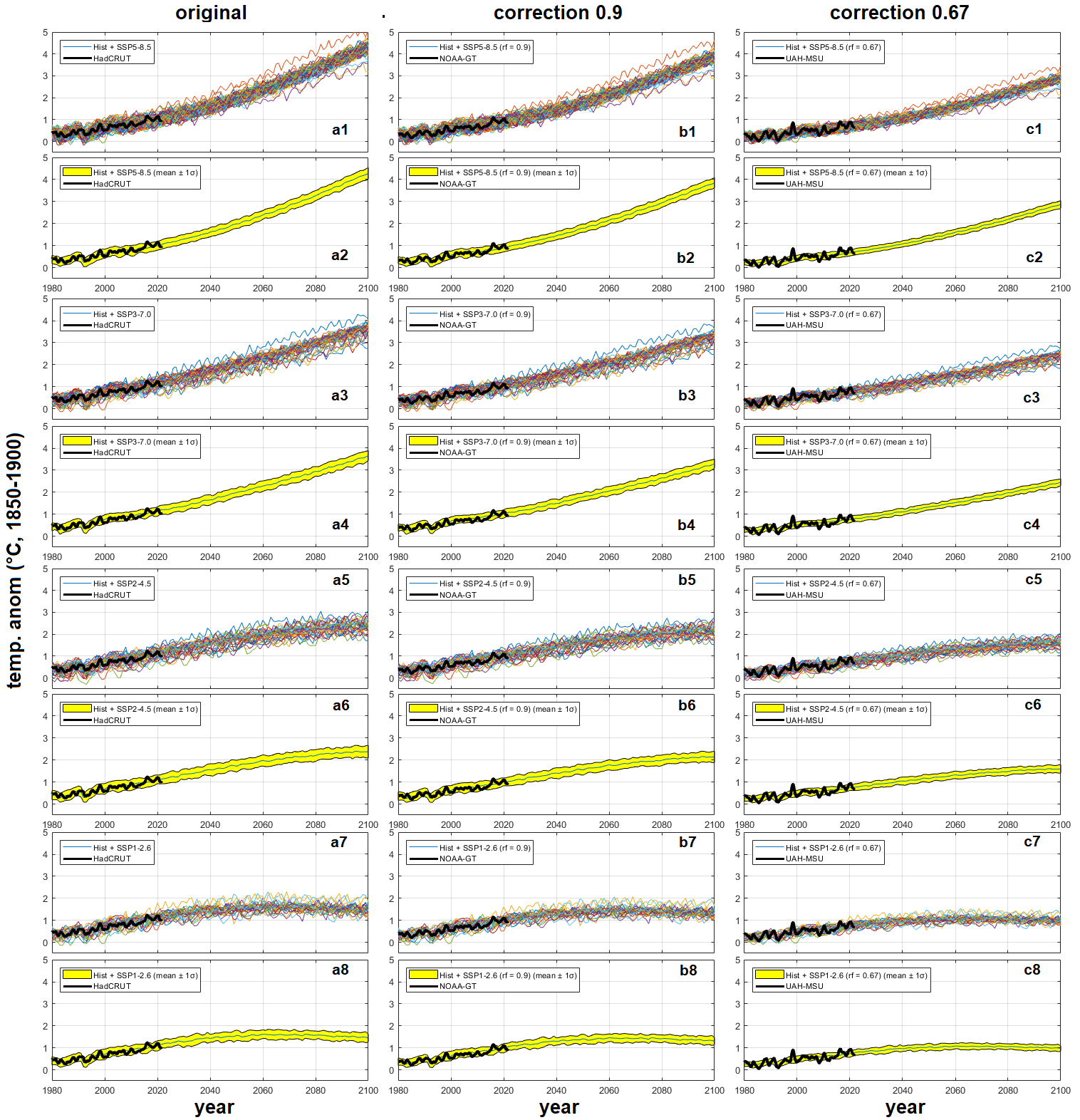}\caption{Low-ECS GCM simulations from 1980 to 2100 using historical + SSP1-2.6,
SSP2-4.5, SSP3-7.0, and SSP5-8.5 scenarios: (a1-a8) original GCM simulations
versus HadCRUT5; (b1-b8) GCM simulations reduced by 10\% versus NOAAGlobalTemp
v5; (c1-c8) GCM simulations reduced by 33\% versus UAH-MSU-lt v6.
The ordinates represent the temperature anomaly relative to the 1850-1900
average of the correspondent model set. The temperature records are
baselined with the GCM simulations in 1980-1990.}
\label{fig9}
\end{figure}

\begin{table}
\centering{}%
\begin{tabular}{llc||cc||cc||cc||c}
﻿ &  & \multicolumn{2}{c}{Hist + SSP1-2.6} & \multicolumn{2}{c}{Hist + SSP2-4.5} & \multicolumn{2}{c}{Hist + SSP3-7.0} & \multicolumn{2}{c}{Hist + SSP5-8.5}\tabularnewline
 &  & \multicolumn{2}{c}{(${^\circ}$C)} & \multicolumn{2}{c}{(${^\circ}$C)} & \multicolumn{2}{c}{(${^\circ}$C)} & \multicolumn{2}{c}{(${^\circ}$C)}\tabularnewline
\hline 
\multirow{4}{*}{Original} & 1980-1990 & \multicolumn{2}{c}{0.43 $\pm$ 0.19} & \multicolumn{2}{c}{0.43 $\pm$ 0.21} & \multicolumn{2}{c}{0.45 $\pm$ 0.20} & \multicolumn{2}{c}{0.37 $\pm$ 0.19}\tabularnewline
 & 2011-2021 & \multicolumn{2}{c}{1.01 $\pm$ 0.20} & \multicolumn{2}{c}{1.01 $\pm$ 0.22} & \multicolumn{2}{c}{1.05 $\pm$ 0.19} & \multicolumn{2}{c}{0.90 $\pm$ 0.21}\tabularnewline
 & 2040-2060 & \multicolumn{2}{c}{1.52 $\pm$ 0.21} & \multicolumn{2}{c}{1.75 $\pm$ 0.25} & \multicolumn{2}{c}{1.97 $\pm$ 0.21} & \multicolumn{2}{c}{1.98 $\pm$ 0.21}\tabularnewline
 & 2080-2100 & \multicolumn{2}{c}{1.52 $\pm$ 0.21} & \multicolumn{2}{c}{2.32 $\pm$ 0.25} & \multicolumn{2}{c}{3.28 $\pm$ 0.24} & \multicolumn{2}{c}{3.78 $\pm$ 0.24}\tabularnewline
\hline 
\multirow{4}{*}{RF = 0.90} & 1980-1990 & \multicolumn{2}{c}{0.39 $\pm$ 0.17} & \multicolumn{2}{c}{0.39 $\pm$ 0.19} & \multicolumn{2}{c}{0.41 $\pm$ 0.18} & \multicolumn{2}{c}{0.33 $\pm$ 0.17}\tabularnewline
 & 2011-2021 & \multicolumn{2}{c}{0.91 $\pm$ 0.18} & \multicolumn{2}{c}{0.91 $\pm$ 0.20} & \multicolumn{2}{c}{0.95 $\pm$ 0.17} & \multicolumn{2}{c}{0.81 $\pm$ 0.19}\tabularnewline
 & 2040-2060 & \multicolumn{2}{c}{1.37 $\pm$ 0.19} & \multicolumn{2}{c}{1.58 $\pm$ 0.23} & \multicolumn{2}{c}{1.77 $\pm$ 0.19} & \multicolumn{2}{c}{1.78 $\pm$ 0.19}\tabularnewline
 & 2080-2100 & \multicolumn{2}{c}{1.37 $\pm$ 0.19} & \multicolumn{2}{c}{2.09 $\pm$ 0.23} & \multicolumn{2}{c}{2.95 $\pm$ 0.22} & \multicolumn{2}{c}{3.40 $\pm$ 0.22}\tabularnewline
\hline 
\multirow{4}{*}{RF = 0.67} & 1980-1990 & \multicolumn{2}{c}{0.29 $\pm$ 0.13} & \multicolumn{2}{c}{0.29 $\pm$ 0.14} & \multicolumn{2}{c}{0.30 $\pm$ 0.13} & \multicolumn{2}{c}{0.25 $\pm$ 0.13}\tabularnewline
 & 2011-2021 & \multicolumn{2}{c}{0.68 $\pm$ 0.13} & \multicolumn{2}{c}{0.68 $\pm$ 0.15} & \multicolumn{2}{c}{0.70 $\pm$ 0.13} & \multicolumn{2}{c}{0.60 $\pm$ 0.14}\tabularnewline
 & 2040-2060 & \multicolumn{2}{c}{1.02 $\pm$ 0.14} & \multicolumn{2}{c}{1.17 $\pm$ 0.17} & \multicolumn{2}{c}{1.32 $\pm$ 0.14} & \multicolumn{2}{c}{1.33 $\pm$ 0.14}\tabularnewline
 & 2080-2100 & \multicolumn{2}{c}{1.02 $\pm$ 0.14} & \multicolumn{2}{c}{1.55 $\pm$ 0.17} & \multicolumn{2}{c}{2.20 $\pm$ 0.16} & \multicolumn{2}{c}{2.53 $\pm$ 0.16}\tabularnewline
\hline 
\end{tabular}\caption{Low-ECS GCMs: global surface warming in the periods 1980-1990, 2011-2021,
2040-2060, and 2080-2100 using the simulations depicted in Figure
\ref{fig9}. Original GCM simulations; (RF = 0.90) GCM simulations
reduced by 10\%; (RF = 0.67) GCM simulations reduced by 33\%. The
temperature anomaly is relative to the 1850-1900 average of the correspondent
model set. The temperature records are baselined with the models simulations
in 1980-1990.}
 \label{tab8}
\end{table}

\section{Climate change expectations for the 21\protect\textsuperscript{st}
century}

Climate impacts several areas of economic and environmental importance
and its changes may require the implementation of various adaptation
policies. However, climate change could also adversely affect some
of the Earth's climate systems such as in areas of water scarcity,
coastal communities, natural ecosystems and others \citep{IPCC2022}.
It is reasonable to assume that if climate change is too rapid and
too significant, different areas could reach a point of vulnerability
where adaptation will no longer be sufficient to avoid serious adverse
effects. However, adaptation policies are much more affordable than
mitigation ones, so the risks associated with possible future climate
change should not be overestimated.

The \citet{IPCC2021} used the GCM CMIP6 and various scenarios of
global socioeconomic change predicted up to 2100 to produce hypothetical
future stories on climate change for the 21\textsuperscript{st} century.
Four SSP scenarios were studied here: the SSP1-2.6 (low GHG emissions
in which CO\textsubscript{2} emissions are reduced to zero around
2075); SSP2-4.5 (intermediate GHG emissions in which CO\textsubscript{2}
emissions increase around the current rate until 2050, and then decrease
but not reach net zero by 2100); SSP3-7.0 (high GHG emissions where
CO\textsubscript{2} emissions double by 2100); and SSP5-8.5 (very
high GHG emissions where CO\textsubscript{2} emissions triple by
2075).

The \citet{IPCC2022} states that if the global surface temperature
rises significantly above 2${^\circ}$C over the next few decades
compared to the pre-industrial period (1850-1900), adaptation policies
may not be sufficient to reduce high risks related to climate change.
Aggressive climate mitigation policies should therefore be implemented
because the CMIP6 GCMs predict that the temperature will likely increase
between 2°C and 3${^\circ}$C (compared to 1850-1900) by 2050 if anthropogenic
greenhouse gas emissions are not significantly reduced as soon as
possible.

However, in the previous sections we found that only the GCM group
with low ECS, which is also the one predicting less warming, optimally
reproduces the observed warming from 1980-1990 to 2011-2021 reported
by the surface-based global temperature records. Therefore, its scenario
forecasts for the 21\textsuperscript{st} century should be preferred
for policy. Furthermore, we also found that global warming from 1980-1990
to 2011-2021 reported by the surface temperature records may need
to be reduced on average by about 10\% assuming that the ocean warming
is correct and that the correct land/ocean temperature ratio is the
one predicted by the models. Finally, if UAH-MSU-lt better reproduces
the actual warming from 1980-1990 to 2011-2021, even the simulations
of the low ECS GCMs would be running too hot and the warming they
produce would need to be reduced by 33\% to optimally accommodate
the observations. Here, we show and discuss how the climate could
change in the 21\textsuperscript{st} century under the above assumptions.

Figure \ref{fig9} shows the simulations produced by the low ECS GCMs
from 1980 to 2100 using the historical + SSP1-2.6, SSP2-4.5, SSP3-7.0,
and SSP5-8.5 scenarios in three conditions: panels a1-a8 show the
original GCM simulations versus HadCRUT5; panels b1-b8 show the GCM
simulations reduced by 10\% compared to NOAAGlobalTemp v5; and panels
c1-c8 show the GCM simulations reduced by 33\% compared to UAH-MSU-lt
v6. The ordinates represent the temperature anomaly relative to the
1850-1900 average of the corresponding GCM set. The temperature records
are baselined with the model simulations in 1980-1990. Table \ref{tab8}
reports the global surface warming forecasts produced by the low ECS
GCMs in the periods 1980-1990, 2011-2021, 2040-2060 and 2080-2100
in the same three conditions.

The analysis shows that the expected warming of the low-ECS GCMs by
2040-2060 is close to 2${^\circ}$C also for the SSP3-7.0 and SSP5-8.5
scenarios, which \citet{Hausfather} described as ``unlikely'' and
as ``highly unlikely'', respectively. However, if the surface temperature
records contain a warming bias and, therefore, the GCM simulations
need to be scaled down to better agree with the actual warming, the
projected warming for 2040-2060 could be lower (or even significantly
lower if UAH -MSU-lt v6 is correct) than 2${^\circ}$C also for the
SSP3-7.0 and SSP5-8.5 scenarios.

There is indirect evidence that the surface-based temperature reconstructions
could be affected by non-climatic warming biases. In fact, compared
to the 1850-1900 mean, the average warming of 1980-1990 is 0.54${^\circ}$C
for HadCRUT5 (infilled data), 0.48${^\circ}$C for GISTEMP v4 (using
the period 1880-1900) and 0.47${^\circ}$C for NOAAGlobalTemp v5 (using
the period 1880-1900). However,  the low ECS GCMs  give a slightly
lower 1980-1990 warming, which is $0.41\pm0.20$ ${^\circ}$C by averaging
all GCM simulations although they better hindcast the 1980-2021 warming.
In contrast, the medium and high ECS GCMs give  $0.48\pm0.27$ ${^\circ}$C
and $0.47\pm0.23$ ${^\circ}$C, respectively, which better fit the
climate records; but then these same GCMs fail to hindcast the observed
warming from 1980 to 2021.

The above findings suggest that climate change will likely be moderate
over the next few decades and that adaptation policies should be sufficient
to manage any adverse effects that may occur.

Furthermore, the warming hindcasted by the low-ECS models from 1850-1900
to 1980-1990 would be lower than $0.41\pm0.20$ ${^\circ}$C if the
climate simulations produced by them were to be scaled down. This
evidence would suggest that the more recently applied homogenization
adjustments to climate data to attempt to remove their non-climatic
biases may have been inadequate and may have added or left spurious
  warming. For example,the continuous homogenization adjustments made
to the surface-based temperature records during the last 10 years
may have improperly cooled the raw temperature data of the past for
many land stations \citep{DAleo(2016)} and, simultaneously, may have
improperly increased the warming trend from the 1970s to the present,
and, in particular, that of the period 2000-2021 \citep{Connolly,Scafetta2021,Watts(2022)}. 

\begin{figure}[!t]
\centering{}\includegraphics[width=1\textwidth]{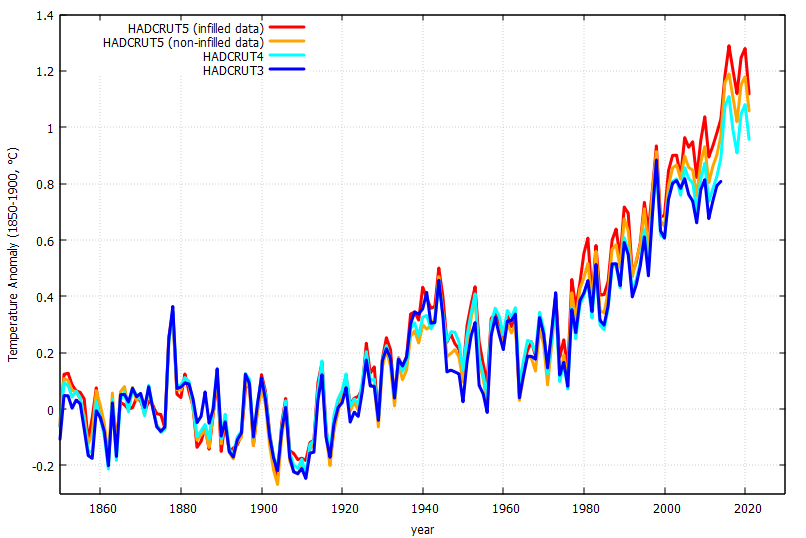}\caption{HadCRUT3 (1850-2014), HadCRUT4 (1850-2021), HadCRUT5 (non-infilled
data, 1850-2021) and HadCRUT5 (filled data, 1850-2021) global surface
temperature records relative to the 1850-1900 period.}
\label{fig10}
\end{figure}

In fact, the scientific literature has indicated the period 2000-2014
as a ``hiatus'' or ``pause'' in global warming \citep{IPCC2013}
because all surface and satellite climate records available before
2014 (e.g. HadCRUT3, which was discontinued in 2014, \citealt{Brohan}) showed
more than a decade of relatively little change. Later, however, new
versions of the surface temperature records were published (e.g. HadCRUT4
and later HadCRUT5 non-infilled and infilled data) and the 2000-2014 “pause”
has gradually disappeared because, from one climate version to the
following one, it has been replaced by an increasingly strong warming
trend; e.g. the trend was 0.03 °C/decade for HadCRUT3, 0.08 °C/decade
for HadCRUT4, 0.10 °C/decade for HadCRUT5 non-infilled data, and 0.14
°C/decade for HadCRUT5 infilled data. See Figure \ref{fig10} and
Table \ref{tab9}. Yet, the 2000-2014 global warming “hiatus” is still
visible in the UAH-MSU-lt v6 record, which shows a 2000-2014 warming
trend of 0.012 °C/decade (Figure \ref{fig2}). 

\begin{table}
\centering{}%
\begin{tabular}{l|cccccc|cc}
﻿ & \multicolumn{6}{c|}{temperature anomaly (°C, 1850-1900)} & \multicolumn{2}{c}{trend (°C/year)}\tabularnewline
 & 1960-1980 & 1980-1990 & 1990-2000 & 2000-2010 & 2004-2014 & 2011-2021 & 2000-2014 & 2000-2021\tabularnewline
\hline 
HadCRUT5 (infilled data) & 0.28 & 0.53 & 0.68 & 0.89 & 0.94 & 1.12 & 0.014 & 0.022\tabularnewline
HadCRUT5 (non infilled data) & 0.24 & 0.49 & 0.65 & 0.83 & 0.87 & 1.04 & 0.010 & 0.019\tabularnewline
HadCRUT4 & 0.25 & 0.42 & 0.59 & 0.78 & 0.81 & 0.95 & 0.008 & 0.016\tabularnewline
HadCRUT3 & 0.24 & 0.43 & 0.58 & 0.76 & 0.76 &  & 0.003 & \tabularnewline
\end{tabular}\caption{Warming means and trends in various periods for HadCRUT3 (1850-2014),
HadCRUT4 (1850-2021), HadCRUT5 (non-infilled data, 1850-2021) and
HadCRUT5 (filled data, 1850-2021) global surface temperature records.
See Figure \ref{fig10}.}
 \label{tab9}
\end{table}

\section{Conclusion}

Here I tested how well the CMIP6 GCMs -- grouped into low, medium
and high ECS subgroups -- hindcast the global surface temperature
warming from 1980-1990 to 2011-2021 reported by four surface temperature
records (ERA5-T2m, HadCRUT5, GISTEMP v4, and NOAAGlobTemp v5) and
by the satellite-based UAH-MSU-lt v6 temperature record. The latter
was used as the lowest possible estimate for the global surface temperature
warming during the analyzed period. The rationale for adding a comparison
with the lower troposphere temperature record is that surface temperatures
could be affected by significant non-climatic warming bias due, for
example, to poorly corrected urban heats and many other factors \citep{Connolly,DAleo(2016),Scafetta2021,Watts(2022)}.
For example, indirect evidence for a significant warming bias, especially over
land, may be also provided by the so-called \textquotedbl Divergence
Problem\textquotedbl{} that is the apparent decoupling between three
ring width chronologies and the rising temperature measurements starting
from the 1970s \citep{Buntgen,Esper,Scafetta2021}.

Using the 143 GCM mean simulations available for four different SSPs,
all medium and high ECS models turn out to be warmer than observations.
Using the 688 CMIP6 ensemble member simulations available, 94\% to
100\% of the simulations produced by GCMs with medium and high ECS
hindcasted greater warming than the five temperature records. In contrast,
the low-ECS models are statistically distributed around the observed
warming values obtained from the four surface-based temperature records.
However, if the UAH-MSU-LT record better represents the actual 2011-2021
warming, even the low-ECS GCM group would produce on average too hot
hindcasts.

I also tested whether the internal variability of the models could
produce results distributed around the observations. Its effect was
modeled using three fixed precision options. Assuming high ($\sigma_{H}\approx0.05$${^\circ}$C)
and medium ($\sigma_{M}\approx0.10$${^\circ}$C) precision, it was
found that 98-100\% and 92 -98\%, respectively, of all possible outputs
from the medium and high ECS GCMs would be warmer than observations.
Only the theoretical results produced by the low-ECS GCM group optimally
 agree  with the surface-based temperature records. If the required
model accuracy is quite low ($\sigma_{L}\approx0.15$ ${^\circ}$C),
the middle and high GCM simulation groups would agree better with
the data, but this agreement could still be quite unsatisfactory because
87\% to 93\% (which is still well outside the $\pm1$$\sigma$ or
68\% confidence interval) of their hindcasts would still be too hot.
In any case, the low precision option should be considered very unsatisfactory
because it would allow the GCMs to deviate too much from the observations.
Moreover, such poor precision would not seem consistent with the natural
variability of the data as argued in Appendix B. Figure \ref{fig5}
suggests that such a low precision could only occur for EC-Earth3
GCM.

Figures \ref{fig5} and \ref{fig6} also show that very few GCMs with
medium and high ECS could produce some simulations consistent with
the actual temperature values. In particular, the two high-ECS CNRM
models \citep{Seferian,Voldoire} appear to perform better than the
other models of the same group. However, as a group, the high-ECS
models are physically incompatible with the low-ECS ones. Indeed,
the internal parameters of the GCMs are carefully tuned to obtain
results as acceptable as possible \citep{Hourdin,Mauritsen}. Therefore,
the good performance of some isolated cases could hardly be used to
validate the corresponding model since the tuning operations also
risk masking fundamental physical problems and, therefore, the need
for model and/or forcing improvements.

It was found that only the low-ECS GCM group agrees optimally with
the surface-based temperature records because their full hindcast
range well encompasses the actual temperature warming values  from
1980-1990 to 2011-2021. Therefore, since the three ECS chosen ranges
 should be considered large enough to be incompatible with each other,
the GCM group with low ECS should be preferred to the other two, implying
that the most likely ECS should be equal to or lower than 3${^\circ}$C.
This result confirms \citet{Scafetta}. In fact, the performance of
the models seems to increase as the ECS decreases \citep{ScafettaMDPI2021}.

However, the actual ECS could also be significantly lower than 3${^\circ}$C
if the UAH-MSU-lt record better represents the 2011-2021 surface warming.
In fact, the satellite record shows that from 1980-1990 and 2011-2021
the global surface temperature may have warmed by about 0.40${^\circ}$C,
which is about 30\% less than 0.58${^\circ}$C as reported by ERA5-T2m,
HadCRUT5, and GISTEMP v4. In this case, even the GCM group with low
ECS would show poor accuracy in reproducing the temperature data because
their average hindcast is about 0.60${^\circ}$C. This means that
the actual ECS could also be 33\% lower than that which characterizes
the low ECS GCM group: that is, it could need to be reduces from 1.8-3.0${^\circ}$C
to 1.2-2.0${^\circ}$C. This conclusion cannot be ruled out because:
(1) the surface temperature records appear to be severely affected
by non-climatic warming bias \citep{Connolly,DAleo(2016),Scafetta2021,Watts(2022)},
as the direct comparison between land and ocean warming proposed here
also seeems to confirm (Figure \ref{fig8}); (2) because a number
of independent studies have concluded that the ECS could be within
such a low range \citep[e.g.:][]{Lewis,Lindzen,Scafetta2013,Stefani,Happer}.

There is a third possibility which would also imply that the actual
ECS should be relatively low. The climate system, in fact, appears
to be also modulated by multidecennial and millennial natural oscillations
such as those related to solar forcings and other astronomical ones,
which are not reproduced by the GCMs \citep[cf.:][]{Scafetta2013,Scafetta2021c,Wyatt}.
Their presence implies that the ECS of GCMs should be at least halved
\citep[cf.:][]{Loehle(2011),Scafetta(2012a),Scafetta2021c} and could
vary approximately between 1.0${^\circ}$C and 2.5${^\circ}$C, as
found by several independent studies \citep[cf.:][]{Lewis,Lindzen,Scafetta2013,Stefani,Happer}.
If so, future climate warming and changes will be moderate and naturally
oscillating \citep{Scafetta2013,Scafetta2021c} and the rate of global
surface warming should likely remain quite low until 2030-2040, when
solar activity is expected to increase again due to its natural multi-decadal
oscillations \citep[and several others]{Scafetta(2012b),ScafettaBianchini(2022)}.

In any case, even remaining within the theoretical framework of the
CMIP6 GCMs, it should be concluded that only the low ECS GCM group
can be considered sufficiently validated by the global surface warming
observed from 1980-1990 to 2011-2021. Therefore, only the 21\textsuperscript{st}
century climate projections produced by the low ECS GCMs should be
used for policy. For decades to come, these models predict more moderate
warming than the GCM groups with medium and high ECS do for similar
greenhouse gas emission scenarios. By 2050, projected warming is expected
to be around 2${^\circ}$C or less even for the worst greenhouse gas
emission scenarios. This moderate warming should not be considered
particularly alarming because the impact and risk assessments  related
to it are considered ``moderate'' assuming even low to no adaptation
\citep{IPCC2022}. Furthermore, as surface-based temperature records
are likely affected by warming biases and are characterized by natural
oscillations that are not reproduced by the CMIP6 models, the global
warming expected for the next few decades may be even more moderate
than predicted by the low-ECS GCMs and could easily fall within a
safe temperature range where climate adaptation policies will suffice.
Therefore, aggressive mitigation policies aimed at rapidly and drastically
reducing GHG emissions in order to avoid a too rapid rise in temperature
do not seem justified, also because their costs seem to outweigh any
realistic benefits \citep[cf.][]{Bezdek}.

\appendix

\begin{figure}[!t]
\centering{}\includegraphics[width=1\textwidth]{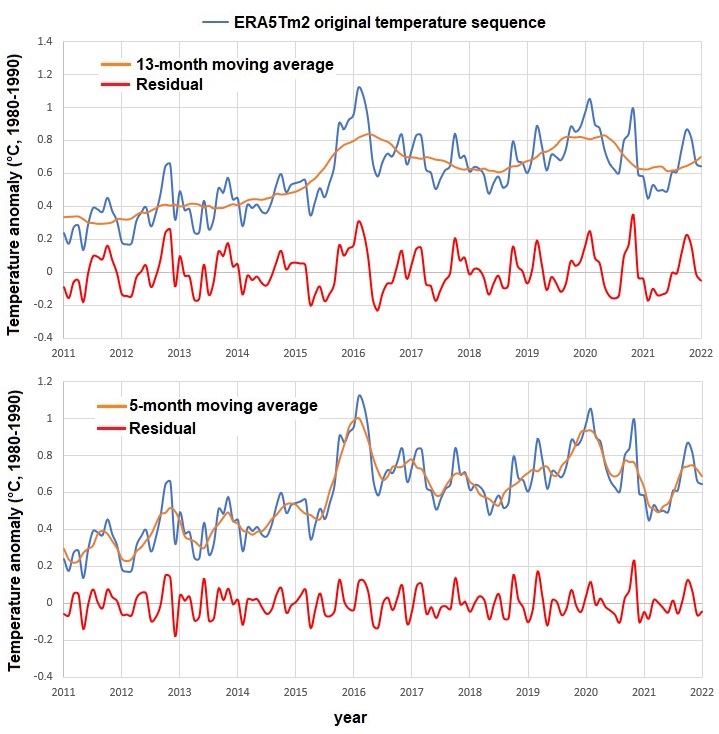}\caption{Possible decomposition of the ERA5-T2m global surface temperature
record (blue) into a signal (orange, made of 13 and 5 month moving
averages) and a residual-noise (red) record that could be used to
evaluate the error of the mean.}
\label{figA1}
\end{figure}

\subsection*{Appendix A: Evaluation of the error of the mean for temperature records}

Computer simulations are made of pure numbers and their averages over
a given period of time are error free. The uncertainty associated
with their unforced internal variability is a different matter and
will be discussed in Appendix B.

Conversely, the data points of the temperature records are affected
by small statistical errors, which however are not always readily
available as is the case with the ERA5-T2m record. Let's address the
issue.

A generic time series $y_{t}$ with $t=1,...,N$ could be affected
by Gaussian distributed uncertainties $\xi_{t}$ with zero mean and
standard deviation $\sigma_{\xi}$ as
\begin{equation}
y_{t}=x_{t}+\xi_{t},\label{eq:2}
\end{equation}
where $x_{t}$ is the physical signal of the record. Its mean is 
\begin{equation}
\bar{y}=\frac{1}{N}\sum_{t=1}^{N}y_{t}\pm\sigma_{\xi}/\sqrt{N},\label{eq:3}
\end{equation}
where $\bar{\sigma}=\sigma_{\xi}/\sqrt{N}$ is the error of the mean.

It is important to note that $\sigma_{\xi}$ is the standard error
of the uncertainties $\xi_{t}$, not that of the signal $y_{t}$.
For example, the ERA5-T2m 2011-2021 average (baselined to 1980-1990)
is 0.578${^\circ}$C, which was obtained by integrating on the globe
the temperature variations that occurred in each cell of the surface
grid worldwide. The standard deviation of the same record is $\sigma_{y,month}=0.20$${^\circ}$C
using the monthly record and $\sigma_{y,year}=0.18$${^\circ}$C using
the annual record. For random variables, the error of the mean does
not depend on the time resolution of the record, that is, the monthly
and yearly resolved records should give $\bar{\sigma}_{year}=\bar{\sigma}_{month}$.
However, if we apply the equation $\bar{\sigma}=\sigma_{y}/\sqrt{N}$
from 2011 to 2021, for ERA5-T2m we get $\bar{\sigma}_{year}=0.054$${^\circ}$C
(using N=11 points) and $\bar{\sigma}_{month}=0.017$${^\circ}$C
(using N=132 points), respectively. This shows that from 2011 to 2021
the ERA-T2m record is not composed of random variables floating around
an average value, but contains a physical signal.

It can be assumed that the physical signal of ERA5-T2m is represented
by the moving averages of the data at 13 months, 5 months or 3 months
while the residuals are the noise components that should be used to
evaluate $\sigma_{\xi}$ and $\bar{\sigma}=\sigma_{\xi}/\sqrt{N}$.
The three choices give $\bar{\sigma}=0.01$${^\circ}$C, $\bar{\sigma}=0.006$${^\circ}$C,
and $\bar{\sigma}=0.005$${^\circ}$C, respectively, which suggest
that the actual error of the 2011-2021 mean could be $\bar{\sigma}=0.01$${^\circ}$C
or probably less. The first two examples of data decomposition are
shown in Figure \ref{figA1}.

Alternatively, the statistical uncertainty associated with ERA5-T2m
could be considered compatible with those explicitly provided by the
other available global surface temperature records. In the case of
the GISTEMP record, \citet{Lenssen} calculated that the resulting
95\% uncertainties are near $\bar{\sigma}_{95\%,annual}\approx0.05$
${^\circ}$C in the global annual mean for the last 50 years. HadCRUT5's
global surface temperature record includes its 95\% confidence interval
estimate and, from 2011 to 2021, the uncertainties for the monthly
and annual averages are $\bar{\sigma}_{95\%,monthly}\approx0.05$${^\circ}$C
and $\bar{\sigma}_{95\%,annual}\approx0.03$${^\circ}$C, respectively.
Berkeley Earth land/ocean temperature record estimates $\bar{\sigma}_{95\%,monthly}\approx0.042$
${^\circ}$C, $\bar{\sigma}_{95\%,annual}\approx0.028$${^\circ}$C,
and $\bar{\sigma}_{95\%,decadal}\approx0.022$${^\circ}$C during
the same period. Note that the error of the mean must decrease as
the time scale increases.

Therefore, adopting the equation $\bar{\sigma}_{95\%}=1.96\times\sigma_{\xi}/\sqrt{N}$,
the probable error for the 2011-2021 mean could be of about 0.01${^\circ}$C
or even smaller. In fact, using the above estimates, we obtain: $0.05/\sqrt{11}=0.015$${^\circ}$C,
$0.05/\sqrt{132}=0.0043$${^\circ}$C, $0.03/\sqrt{11}=0.009$${^\circ}$C,
$0.042/\sqrt{132}=0.004$${^\circ}$C, and $0.028/\sqrt{11}=0.008$${^\circ}$C
respectively, the mean of which is approximately 0.008${^\circ}$C.
Alternatively, the 95\% uncertainty over the period 2011-2021 cannot
be greater than about $\pm$0.02${^\circ}$C, as explicitly reported
by the Berkeley Earth land/ocean temperature record for the ten-year
scale.

Therefore, various methodologies suggest that the  uncertainty of
the temperature means in the 11-year period from 2011 to 2021 is very
small, around $\pm0.01$${^\circ}$C at 95\% confidence and can be
safely ignored as done for example in \citet{Scafetta}.

\begin{figure}[!t]
\centering{}\includegraphics[width=1\textwidth]{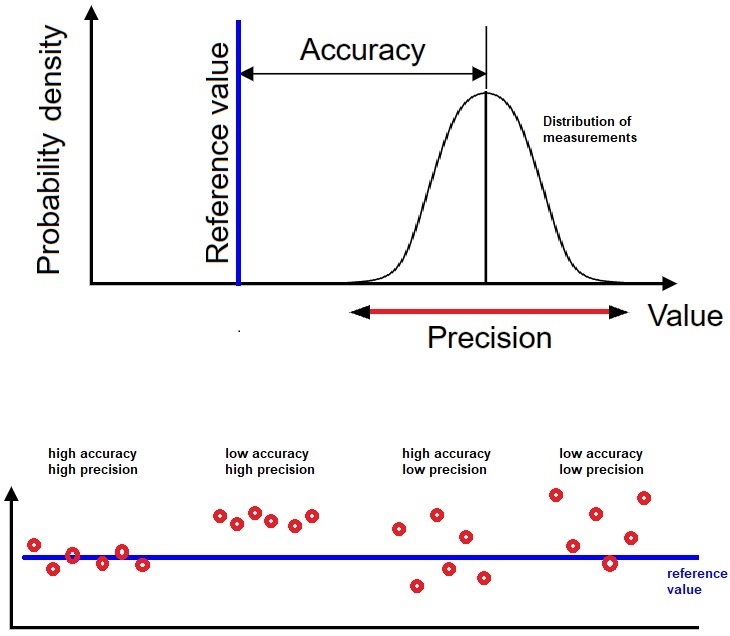}\caption{Definition of accuracy and precision, and examples of high and low
accuracy and precision cases where the red dots are ensembles of measurements.}
\label{figA2}
\end{figure}

\begin{figure}[!t]
\centering{}\includegraphics[width=1\textwidth]{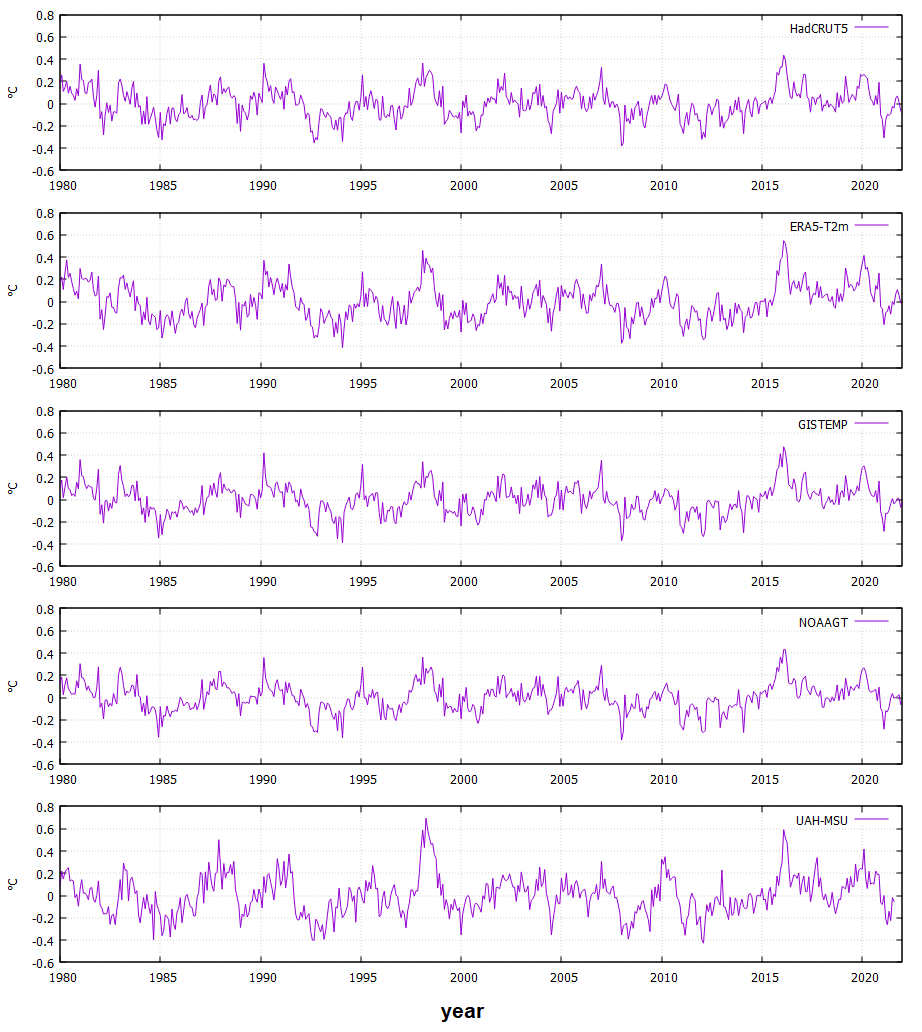}\caption{15-year high-pass filter of HadCRUT5, ERA5-T2m, GISTEMP v4, NOAAGlobTemp
v5, and UAH-MSU-lt v6 temperature records.}
\label{figA3}
\end{figure}

\subsection*{Appendix B: Provisional assessment of an acceptable internal variability
of the models}

Climate models are unable to accurately simulate temperature records
due to various uncertainties. For example, only by varying the initial
conditions different climatic trajectories are obtained which evolve
around an ideal average with a certain variance.

Each GCM is characterized by its own internal variability. However,
in the main text, we argued that such uncertainty could not be arbitrarily
large because the models must be both accurate and precise. Figure
\ref{figA2} explains the concepts of “accuracy”, which measures the
proximity of the model hindcasts to the real value, and of “precision”,
which indicates the proximity of the same hindcasts to each other.

An acceptable range for the distribution of hindcasts related to the
internal variability of the models could be assessed by observing
that temperature fluctuations at time scales lower than, for example,
15 years, (which include the climatic oscillations of the ENSO and
those of the 11-year solar cycle) may not be adequately predicted
by the models. From the point of view of the models, the temperature
fluctuations at those time scales could be considered stochastic and
their standard deviation from the mean could represent the acceptable
range that can be associated with the unforced internal variability
of the models \citep[cf.][]{Knight}.

Figure \ref{figA3} shows the 15-year high-pass filter curves of HadCRUT5,
ERA5-T2m, GISTEMP v4, NOAAGlobTemp v5, and UAH-MSU-lt v6 temperature
records from 1980 to 2021. Using the standard deviation $\sigma$
for each record, the error of the mean on 11-year intervals (e.g.
from 2011 to 2021) at the 95\% confidence is $\bar{\sigma}_{95\%}=1.96\times\sigma/\sqrt{11}$,
that is: $\bar{\sigma}_{95\%}=0.080$${^\circ}$C for HadCRUT5; $\bar{\sigma}_{95\%}=0.094$${^\circ}$C
for ERA5-T2m; $\bar{\sigma}_{95\%}=0.079$${^\circ}$C for GISTEMP
v4; $\bar{\sigma}_{95\%}=0.075$${^\circ}$C for GISTEMP v4; and $\bar{\sigma}_{95\%}=0.104$${^\circ}$C
for UAH-MSU-lt v6.

Based on the above assumptions, over an 11-year period, the uncertainty
of the 2011-2021 warming compared to 1980-1990 could be estimated
at approximately $\pm0.1$${^\circ}$C at the 95\% confidence, which
corresponds to the high precision option ($\sigma_{H}=0.05$${^\circ}$C)
discussed in Section 3.

The high precision option should not be interpreted as the actual
dispersion produced by each GCM, which varies greatly from model to
model, but only as the acceptable uncertainty that a CGM should exhibit
in reproducing the warming from 1980-1990 to 2011-2021. In our case,
a $\pm0.1$${^\circ}$C error would imply a $\pm0.17\%$ of the actual
warming from 1980 to 2021, which can be considered a reasonable error.

\subsection*{Conflict of Interest}

The author declares no conflict of interest.

\subsection*{Data Availability Statement}

All data used in the manuscript can be downloaded from KNMI Climate
Explorer or from the original websites:
\begin{itemize}
\item KNMI Climate Explorer, \href{https://climexp.knmi.nl/start.cgi}{https://climexp.knmi.nl/start.cgi}
\item ERA5-T2m, \href{https://cds.climate.copernicus.eu/cdsapp\#!/dataset/reanalysis-era5-single-levels-monthly-means}{https://cds.climate.copernicus.eu/cdsapp\#!/dataset/reanalysis-era5-single-levels-monthly-means}
\item GISTEMP v4, \href{https://data.giss.nasa.gov/gistemp}{https://data.giss.nasa.gov/gistemp},
\href{https://data.giss.nasa.gov/gistemp/graphs_v4/}{https://data.giss.nasa.gov/gistemp/graphs\_v4/}
\item HadCRUT5, \href{https://www.metoffice.gov.uk/hadobs/hadcrut5/}{https://www.metoffice.gov.uk/hadobs/hadcrut5/}
\item HadCRUT4, \href{https://www.metoffice.gov.uk/hadobs/hadcrut4/}{https://www.metoffice.gov.uk/hadobs/hadcrut4/}
\item HadCRUT3, \href{https://www.metoffice.gov.uk/hadobs/hadcrut3/}{https://www.metoffice.gov.uk/hadobs/hadcrut3/}
\item NOAAGlobTemp v5, \href{https://www.eea.europa.eu/data-and-maps/data/external/noaa-global-temperature-v5}{https://www.eea.europa.eu/data-and-maps/data/external/noaa-global-temperature-v5}
\item Berkeley Earth land/ocean temperature, \href{http://berkeleyearth.org/archive/land-and-ocean-data/}{http://berkeleyearth.org/archive/land-and-ocean-data/}
\item Japanese Meteorological Agency, \href{https://ds.data.jma.go.jp/tcc/tcc/products/gwp/temp/ann_wld.html}{https://ds.data.jma.go.jp/tcc/tcc/products/gwp/temp/ann\_wld.html}
\item UAH-MSU-lt v6, \href{https://www.nsstc.uah.edu/data/msu/v6.0/tlt/uahncdc_lt_6.0.txt}{https://www.nsstc.uah.edu/data/msu/v6.0/tlt/uahncdc\_lt\_6.0.txt}
\item CMIP6 GCMs, \href{http://climexp.knmi.nl/selectfield_cmip6.cgi?id=someone@somewhere}{http://climexp.knmi.nl/selectfield\_cmip6.cgi?id=someone@somewhere}.
\end{itemize}

\end{document}